**Title**

Mesosiderite formation on asteroid 4 Vesta by a hit-and-run collision

**Authors**

Makiko K. Haba[1,2*], Jörn-Frederik Wotzlaw[1], Yi-Jen Lai[1†], Akira Yamaguchi[3] and Maria Schönbächler[1]

**Affiliations**

[1]ETH Zürich, Institute of Geochemistry and Petrology, 8092 Zürich, Switzerland.

[2]Department of Earth and Planetary Sciences, Tokyo Institute of Technology, Ookayama, Tokyo 152-8551, Japan.

[3]National Institute of Polar Research, Tachikawa, Tokyo 190-8518, Japan.

[*]Corresponding author. E-mail: haba.m.aa@m.titech.ac.jp.

[†]Present address: Macquarie GeoAnalytical, Department of Earth and Planetary Sciences, Macquarie University, 12 Wally's Walk, Sydney, New South Wales 2109, Australia.

**Abstract**

Collision and disruption processes of proto-planetary bodies in the early solar system are key to understanding the genesis of diverse types of main-belt asteroids. Mesosiderites are stony-iron meteorites that formed by mixing of howardite-eucrite-diogenite-like crust and molten core materials and provide unique insights into the catastrophic break-up of differentiated asteroids. However, the enigmatic formation process and the poorly constrained timing of metal-silicate mixing complicate the assignment to potential parent bodies. Here we report high-precision uranium-lead dating of mesosiderite zircons by isotope dilution thermal ionization mass spectrometry, revealing initial crust formation 4,558.5 ± 2.1 million years ago and metal-silicate mixing at 4,525.39 ± 0.85 million years. The two distinct ages coincide with the timing of crust formation and a large-scale reheating event on the eucrite parent body, likely the asteroid Vesta. This chronological coincidence corroborates that Vesta is the parent body of mesosiderite silicates. Mesosiderite formation on Vesta can be explained by a hit-and-run collision 4,525.4 million years ago that caused the thick crust observed by NASA's Dawn mission and explains the missing olivine in mesosiderites, howardite-eucrite-diogenite meteorites, and vestoids.

The accretion, differentiation, and collisional disruption history of asteroidal bodies is important for understanding the formation of protoplanets and diverse types of main-belt asteroids. Differentiated meteorites provide unique records of the timing and duration of these processes. Mesosiderites are differentiated meteorites consisting of almost equal amounts of silicates and Fe-Ni metal. The geochemical characteristics of mesosiderites imply a formation process that requires contributions from crust and core materials without mantle material[1–3]. Thus, mesosiderite formation involves the catastrophic disruption of at least one differentiated asteroid. The petrology and chemical compositions of mesosiderite clasts are similar to howardite-eucrite-diogenite (HED) meteorites; particularly with howardites, which are brecciated mixtures of eucrites and diogenites[4]. Oxygen and Cr isotope compositions of mesosiderite silicates are consistent with those of HED meteorites[5,6] and $^{53}$Mn-$^{53}$Cr systematics of mesosiderites, eucrites, and diogenites define a single isochron line, indicating that their parent bodies simultaneously experienced global silicate differentiation 2.4 million years (Myr) after the formation of calcium-aluminium-rich inclusions[7]. Since HED meteorites are believed to originate from asteroid 4 Vesta[8–10], the linkages with mesosiderite silicate imply that mesosiderites also came from Vesta or that their parent bodies formed in neighbouring regions.

Mesosiderite metal displays a chondritic siderophile element composition with limited variability suggesting that it originated from the metallic core of a differentiated asteroid that was largely molten during mixing with the silicates[2]. Previously proposed formation processes either consider the metal originating from the same parent body as the mesosiderite silicates[3] (internal origin) or deriving from the core of a different asteroid that collided with the silicate parent body[1,2] (external origin). However, the formation process and respective parent bodies are still poorly constrained. Importantly, the timing of the metal-silicate mixing event that formed mesosiderites has not been determined, although a lower age limit of 100–150 Myr after solar system formation has been inferred[11,12]. In either formation model, Vesta as inferred silicate parent body would have experienced a significant impact associated with reheating that might have left a chronological mark on both mesosiderites and HED meteorites.

Zircon [$ZrSiO_4$] is a rare mineral in asteroidal meteorites, but was identified in mesosiderites[13–15]. In situ $^{207}$Pb-$^{206}$Pb and Hf-W dates from a total of five zircons reveal a complex history with crystallization ages ranging from 4,563 Myr (ref. 13) to ~4,520 Myr (refs 14,15). The older dates were interpreted to correspond to the time of crust formation on the silicate parent body. The younger zircon dates indicate that the silicate parent body experienced a later reheating event. However, the small dataset and low precision of individual dates prevented resolving the cause, mode, and duration of the reheating event as well as the temporal relationship with thermal events on potential parent bodies.

Here we report the first comprehensive high-precision U-Pb geochronology dataset for zircons from mesosiderites to determine the precise timing of the mixing event. Twenty-four single zircons of ~40–180 μm in diameter and two aliquots that include 2–3 small zircons (<40–70 μm) were hand-picked from acid residues of five mesosiderites (Vaca Muerta, NWA 1242, NWA 8402, Estherville, and NWA 8741). These zircons were analysed for U and Pb isotopes using isotope dilution thermal ionization mass spectrometry (ID-TIMS; see Methods).

**High-precision $^{207}$Pb-$^{206}$Pb dates for mesosiderite zircons**

The $^{207}$Pb-$^{206}$Pb dates of analysed zircons show two distinct populations (Fig. 1,

Supplementary Table 1). Two aliquots from Estherville and Vaca Muerta and one zircon from Estherville yielded older dates with a weighted mean of 4,558.51 ± 2.1 Myr (95% confidence interval including overdispersion[16]). The remaining twenty-three zircons yielded distinctly younger $^{207}$Pb-$^{206}$Pb dates with a weighted mean of 4,525.39 ± 0.85 Myr. The older dates are in good agreement but significantly more precise than the previously reported $^{207}$Pb-$^{206}$Pb date for a magmatic zircon from Vaca Muerta dated by ion microprobe (4,563 ± 15 Myr; ref. 13). These zircons are considered to be relict zircons that formed in the silicates before the metal-silicate mixing event. The weighted mean age of the younger zircon population is indistinguishable but significantly more precise than recently reported $^{207}$Pb-$^{206}$Pb dates of two zircons from Estherville (4,521 ± 26 Myr) (ref. 14) and the $^{182}$Hf-$^{182}$W date (4,532.8 +11.4/-21.0 Myr) (ref. 15) of one zircon from Asuka 882023. In previous studies, the younger ages were proposed to record either the timing of the metal-silicate mixing[14] or later local impacts[14,15].

Mesosiderites are subgrouped into four types based on the recrystallization degree of the silicate matrices and include those with poorly recrystallized texture (type 1), moderately to highly recrystallized texture (types 2 and 3), and melt-rock matrix (type 4) (ref. 17). Our samples cover various degrees of recrystallization: Vaca Muerta is type 1, NWA 1242 type 2, NWA 8402 type 3, Estherville type 3–4, and NWA 8741 type 4. The younger zircon population is present in all mesosiderites with single crystal ages largely overlapping within analytical uncertainty (Fig. 1). Their ages therefore likely record a major global reheating event on the mesosiderite silicate parent body. The younger zircons from previous works are characterized by low U and Th concentrations of <1 ppm (refs 14,15) while the older zircons have U and Th concentrations of 53 and 16 ppm, respectively[13]. Uranium concentrations of fourteen zircons from NWA 8402, Estherville, and NWA 8741 range from 0.1 to 1.2 ppm and model Th/U ratios determined by TIMS range from 0.002 to 0.12, similar to previously reported compositions of younger zircons. Because meteoritic zircons with such low trace element concentrations were solely found in mesosiderites, zircon formation (younger zircon population) in mesosiderites is likely related to a unique thermal event that affected the mesosiderite silicate parent body. Zircon size increases with recrystallization degree of the sample[14] (Supplementary Table 1). Since recrystallization is related to metamorphism during mixing with molten metal[18], we attribute zircon formation to the metal-silicate mixing. Low trace element concentrations are due to zircon crystallization from a melt after incorporation of abundant U, Th, and rare earth elements (REE) into phosphate minerals as a consequence of metal-silicate mixing[14]. Phosphates in mesosiderites are abundant[19] (0.1–3.7 vol.%) when compared to HED meteorites[20] (<0.1–0.4 vol%) and likely formed during metal-silicate mixing by a reaction of Ca from silicates with P from Fe-Ni metal[21]. In agreement with previous works, abundant phosphates were observed in all mesosiderites investigated here (Supplementary Fig. 1). The low trace element concentrations of the younger zircons therefore strongly support that these zircons grew during the metal-silicate mixing. Unlike HED meteorites, gabbroic clasts in mesosiderites are highly depleted in incompatible elements with high Eu/Sm ratios[22]. This feature is likely related to missing phosphates and zircons in the bulk analyses. These minerals are the main hosts for REE, Zr, and Hf and they are often heterogeneously distributed and located along the boundaries between silicates and metal (see Supplementary Information).

Since the mesosiderite metal was in a molten state at the time of mixing, the parent body size can be roughly estimated by comparison with numerical calculations of Vesta's thermo-chemical evolution. The largely molten core materials present at the time of metal-silicate mixing entails that the parent body of mesosiderite metal was a protoplanet ≥530 km in diameter (see Supplementary Information).

**Chronologic fingerprint of the HED parent-body**

Most zircons in basaltic eucrites yield crystallization ages of 4,550–4,560 Myr (refs 23–25) with a weighted average of 4,554.4 ± 1.7 Myr (ref. 25). These dates overlap with the $^{207}$Pb-$^{206}$Pb dates of the older zircons in mesosiderites (Fig. 2a,b), indicating that crustal formation of the parent bodies of mesosiderite silicate and HED meteorites occurred simultaneously. Zircons from the basaltic eucrite Camel Donga yielded a younger weighted mean $^{207}$Pb-$^{206}$Pb date of 4,530 ± 10 Myr (ref. 24). Zircons with $^{207}$Pb-$^{206}$Pb or $^{182}$Hf-$^{182}$W dates around 4,530 Myr are also reported from Millbillillie, Elephant Moraine 90020, Yamato 791438, and NWA 5073 (Supplementary Table 2) (refs 26–29). These younger ages are within the range of the younger zircons from mesosiderites (4,525.39 ± 0.85 Myr). Among the younger zircons in basaltic eucrites, those in Millbillillie and NWA 5073 show heterogeneous texture in cathodoluminescence images with possible younger overgrowth[26,30]. The younger zircons suggest that they experienced high-temperature reheating after initial magmatism or thermal metamorphism. Zircons resolvably younger than 4,530 Myr are not reported for basaltic eucrites, but some radiometric dates using *whole rock and minerals* (e.g., $^{146,147}$Sm-$^{142,143}$Nd, U-Pb, and $^{244}$Pu-Xe dates) record later local impacts (Fig. 2c,d,e). This implies that zircon formation or resetting of the U-Pb and Hf-W systems in zircons do not occur during short-term reheating by local impacts. As postulated for the mesosiderite parent body above, this provides evidence for a prolonged reheating event ~4,530 Myr ago on the HED parent body. Judging from the crystallization temperature of eucritic zircon (~900°C) (ref. 25) and the closure temperature of the U-Pb decay system in zircon (>950°C) (ref. 23), the peak temperature during the reheating event ≥900–950°C.

This division into two major events around 4,550–4,560 Myr and ~4,530 Myr, respectively, is also supported by the distribution of *whole rock data* (Fig. 2c,d,e). The $^{146,147}$Sm-$^{142,143}$Nd, U-Pb, and $^{244}$Pu-Xe chronometers have high closure temperatures (~800–1200°C), but show slightly different resistance to reheating, with the resistance decreasing in the order of Sm-Nd, $^{244}$Pu-Xe, and U-Pb (refs 31,32). The Sm-Nd dates of basaltic eucrites show a mode at 4,550–4,560 Myr (Fig. 2c), recording the closure time during initial magmatism or prolonged initial thermal heating. The distribution of $^{207}$Pb-$^{206}$Pb dates, however, peak around 4,520–4,530 Myr, rather than the crustal formation age (Fig. 2d,e). The different distributions of $^{207}$Pb-$^{206}$Pb and $^{244}$Pu-Xe ages likely reflect their different resistances against reheating and large uncertainties of $^{244}$Pu-Xe dates compared with the U-Pb dates. Because the peak of $^{207}$Pb-$^{206}$Pb dates is similar to the younger dates of eucritic zircons, it is probable that these date the same reheating event. Moreover, the $^{207}$Pb-$^{206}$Pb and $^{244}$Pu-Xe data are from 13 basaltic eucrites with diverse metamorphic grades and range from 4,510 to 4,540 Myr (Supplementary Table 2), indicating that the reheating event affected widespread regions in the eucritic crust. These younger dates likely reflect a reheating event caused by an external heat source, i.e. a large-scale impact, rather than an internal heat source. This is because temperatures of ~900–1,000°C are needed for zircon formation in basaltic eucrites[25,33]. An internal heat source producing such high temperatures no longer existed at ~4,530 Myr because most of the short-lived $^{26}$Al and $^{60}$Fe had already decayed and the produced heat was insufficient to reach such temperatures. Given that the mean age of the large-scale reheating event is in good agreement with those of younger mesosideritic zircons (Fig. 2a), the event is likely caused by the same large-scale impact that lead to metal-silicate mixing on the mesosiderite parent body. These chronological coincidences together with petrological and geochemical similarities strongly indicate that mesosiderites originated from the HED parent body i.e. likely Vesta.

**Formation models of mesosiderites**

The *external-origin model* for mesosiderites assumes that two distinct differentiated asteroids collided at low relative velocity (~1 km/s) to yield a projectile to target material ratio of approximately unity[1,2]. However, it has been argued that the mean impact velocity for main belt asteroids is ~5 km/s, which normally results in almost complete loss of projectile material during collisions[3]. The parent body of mesosiderite metal must have been approximately as big as Vesta to maintain a molten core (see Supplementary Information) and a direct collision between such large protoplanets at a velocity of ~5 km/s likely leads to wholescale disruption[34]. To leave Vesta relatively intact, a sequential collision process is needed: the parent body of the mesosiderite metal was first disrupted by a collision with another planetesimal, and straightway the excavated molten core materials collided with Vesta's surface at a low velocity. In this process, the molten core material ejected from the disrupted protoplanet is widely scattered into space[34] and cools rapidly. To avoid metal solidification before mixing with silicates, the catastrophic collision must occur very close to Vesta. The collision with the debris from the first collision near Vesta results in a large-scale break-up of Vesta's crust and possibly mantle, which globally mixes with collisional debris composed of crust, mantle, and core materials. Since mesosiderite silicates yield identical chemical, mineralogical, and isotopic data to those of HED meteorites[4–6], this model is unlikely, unless a protoplanet with the very same composition as Vesta provided the collisional debris.

As an alternative, the *internal-origin model* proposes to mix the upper crust and core materials of Vesta by a large-scale break-up (Fig. 3). We suggest that this break-up involved a hit-and-run collision with a smaller planetesimal (mass ratio 0.1) at a mean impact velocity of ~5 km/s (refs 34,35). Such a collision will disrupt one hemisphere of Vesta and produces significant ejecta consisting mainly of crust and mantle materials, but also small amounts of molten core[35]. Most of the escaped materials reaccrete on the opposite surface of the collision site on Vesta within a few hours with most core materials still molten, thereby thickening the original crust[34]. This internal-origin model adopting a hit-and-run collision can explain the mesosiderite formation on Vesta and also the following unsolved problems concerning Vesta.

**Evolution of Vesta's crust and south pole region**

Global mapping with high resolution imagery during the Dawn mission was performed on the southern hemisphere composed of two overlapping impact basins, Veneneia and Rheasilvia, to explore Vesta's internal structure[36,37]. This work revealed that the crater insides are largely covered by howardite-like materials, but do not contain olivine-rich rocks. Based on impact simulations, the crust-mantle boundary of Vesta was estimated to deeper than 80 kilometers in that area[38]. This estimate is much thicker than that obtained from magma-ocean crystallization models using the chemical compositions of HED meteorites[39,40]. These predict a crust composed of a 18–22 km-thick eucritic layer overlying a 13–40 km-thick diogenitic layer. Although the simple magma ocean model for diogenites is still debated[41], the maximum crust estimation is not affected. Achieving a crustal thickness of >80 km requires chondritic source material of more than twice the total mass of current Vesta, when considering the REE concentrations of basaltic eucrites (7–10×CI chondrite), which occupy two-thirds of the Vesta's crust[42]. In order to reduce the primordial mass of Vesta, hit-and-run collisions were considered in the very early stage after crust formation[42]. However, a hit-and-run collision generally removes crust rather than mantle from the involved bodies[34] and cannot explain the significant mass loss of Vesta without stripping the crust. In our internal-origin model, the eucritic crustal thickness of primordial Vesta is assumed to be 18–22 km in accordance with magma-ocean crystallization models[40], thereby circumventing the need of significant mass loss. The thick crust observed at Vesta's south pole in the Dawn mission inherently results from the accretion of the hit-and-run ejecta (Fig. 3). The collision must have occurred in the northern hemisphere, where the crust was largely stripped and the exposed mantle is now overlain with regolith derived from the hit-and-run collision and subsequent impacts.

After the hit-and-run collision, Vesta's original crust (Fig. 4a) was covered with accreted materials composed of howarditic and mantle materials (Fig. 4b). In the south pole region, the hit-and-run debris mixed with the underlying crust in the early stage of reaccretion, whereas subsequently pure ejecta debris accumulated in the upper layer. This process produces roughly three layers with >80 km thickness, composed of hit-and-run debris, debris mixed with underlying crust, and original crust overlying the mantle (Fig. 4c). The total thickness of the two top layers probably exceeded at least ~20 km. Considering the extremely slow cooling rate of mesosiderites at temperature below 500°C (~0.4°C/Myr) (refs 43,44) and their low olivine contents[19] (≤6%), they presumably formed in the lower part of the second layer (mixture of debris and underlying crust). Numerous impacts in the aftermath of the hit-and-run collision probably led to metal fragmentation observed in mesosiderites[3]. Much after the hit-and-run collision, the materials in the south pole region were ejected by the Veneneia and Rheasilvia impacts 2.1 ± 0.2 and 1.0 ± 0.2 billion years ago, respectively[45]. According to estimates of the initial depth of the escaped rocks, the Veneneia and Rheasilvia impacts ejected materials from shallow (<25 km) and greater depth (20–80 km), respectively[38]. After the impacts, the ejecta covered the whole Vestan surface except for the center of the south pole basins[46]. Although mantle materials could have been excavated in the northern hemisphere by local impacts after the hit-and-run collision, the current Vestan surface should be overlain with the ejecta from the Rheasilvia basin, originating from deeper layers, enriched in crustal materials and depleted in olivine-rich rocks (Fig. 4c). In addition, vestoids may derive directly from the south pole basins[47] and because they generally strongly erode in space[48], the vestoids that we observe today should be enriched in the ejecta from the younger Rheasilvia impact rather than those from the Veneneia impact. The HED meteorites and mesosiderites may represent this material in our collections and consistent with our model, they lack olivine-rich rocks.

**Methods**

Five mesosiderites, Vaca Muerta (9.5 g), NWA 1242 (12.5 g), NWA 8402 (7.5 g), Estherville (11.9 g), and NWA 8741 (16.4 g) were used in this study. Vaca Muerta and NWA 1242 had fusion crusts on their surfaces, which were carefully removed using a dental drill. Zircons were hand-picked from the mesosiderite samples by dissolving the metal parts in concentrated HCl at 120°C for an hour and the silicate parts in concentrated $HNO_3$-HF mixture at 150°C for 3 hours. Number and size of analyzed zircons from each sample are shown in Supplementary Table 1. Zircon grains were cleaned with ethanol in an ultrasonic bath. Selected crystals were placed into 3 ml Savillex beakers, fluxed in 6

M HCl at 80°C and ultrasonically cleaned in 3 M $HNO_3$. Cleaned crystals were transferred into 200 µl Savillex microcapsules with a microdrop of 7 M $HNO_3$ and 60 µl of 29 M HF. Zircon samples were spiked with 3 to 7 mg of EARTHTIME $^{202}$Pb-$^{205}$Pb-$^{233}$U-$^{235}$U tracer solution [ET2535 (ref. 49)], the microcapsules were assembled in Parr bombs and zircons were dissolved at 210°C for 60 hours. Dissolved samples were dried down and redissolved in 6 M HCl at 180°C in Parr bombs to convert fluorides to chlorides. Samples were dried down again and dissolved in 3 M HCl for anion exchange chromatography. Uranium and lead were separated employing a single-column HCl-based chemistry modified from ref. 50. Both, the U-Pb fractions and the Zr+Hf+REE fractions were collected in precleaned 7 ml Savillex beakers. The U-Pb fractions were dried down with a microdrop of 0.02 M $H_3PO_4$ while the Zr+Hf+REE fractions were dried down without $H_3PO_4$. U-Pb fractions were loaded with 1 µl Si-Gel emitter[51] on outgassed single Re-Filaments and U-Pb isotopic ratios were measured employing a Thermo TRITON plus thermal ionization mass spectrometer at ETH Zurich using analytical protocols documented by refs 52,53. Pb was measured by peak-hopping on the axial secondary electron multiplier. Measured isotopic ratios were corrected online for instrumental mass fractionation using the double spike. U was measured as $UO_2$ using a static collection routine with Faraday cups connected to amplifiers with $10^{13}$ ohm resistors[53]. Isobaric interferences from minor $UO_2$ isotopologues were corrected using an $^{18}O/^{16}O$ of 0.002064 ± 0.000011 [2 SD (ref. 53)]. Instrumental mass fractionation was corrected using the double spike. U blank was estimated to be 0.04 ± 0.02 pg (2 SD) based on four total procedural blank measurements. Due to the low U content of mesosiderite zircons, the uncertainty on the U blank mass is the largest source of uncertainty of U-Pb ratios rendering U-Pb dates imprecise. This does not allow to confidently assess concordance even though all analyses are concordant within uncertainty (Supplementary Fig. 6). Post-crystallization Pb-loss is unlikely considering the low levels of radiation damage as demonstrated by sharp Raman peaks and strong CL emission (Supplementary Figs. 3 and 4). All common Pb in zircon analyses was assumed to be procedural blank and was corrected using the average of twenty-four total procedural blank measurements analyzed over the course of this study that yielded the following average composition (uncertainties are 2 SD): $^{206}Pb/^{204}Pb = 18.41 \pm 0.40$; $^{207}Pb/^{204}Pb = 15.19 \pm 0.39$; $^{208}Pb/^{204}Pb = 36.93 \pm 0.91$. Data reduction, age calculation and uncertainty propagation were performed using the Tripoli and ET_Redux software package[54] that uses algorithms of ref. 55. U-Pb ratios and dates were calculated relative to the published composition of the ET2535 tracer [v. 3.0 (ref. 49)] and the $^{238}U$ and $^{235}U$ decay constants of ref. 56 (Supplementary Table 1). $^{207}Pb/^{206}Pb$ dates were calculated using the decay constants of ref. 56 and a $^{238}U/^{235}U$ ratio of $137.784 \pm 0.022$, based on previously reported bulk U-isotope analyses of inner solar system materials (eucrites, ordinary and carbonaceous chondrites, enstatite chondrites, basaltic angrites[57,58]). U concentration of analyzed zircons from Estherville was calculated from the measured U amount by ID-TIMS and the zircon weight estimated from crystal size as spherical shape and density (4.7 g/cm$^3$) (Supplementary Table 1).

**References**

1. Wasson, J. T. & Rubin, A. E. Formation of mesosiderites by low-velocity impacts: A natural consequence of planet formation. *Nature* **318**, 168–170 (1985).
2. Hassanzadeh, J., Rubin, A. E. & Wasson, J. T. Compositions of large metal nodules in mesosiderites: Links to iron meteorite groups IIIAB and the origin of mesosiderite subgroups. *Geochim. Cosmochim. Acta* **54**, 3197–3208 (1990).

3. Scott, E. R. D., Haack, H. & Love, S. G. Formation of mesosiderites by fragmentation and reaccretion of a large differentiated asteroid. *Meteorit. Planet. Sci.* **36**, 869–881 (2001).

4. Mittlefehldt, D. W., Chou, C.-L. & Wasson, J. T. Mesosiderites and howardites: Igneous formation and possible genetic relationships. *Geochim. Cosmochim. Acta* **43**, 673–688 (1979).

5. Greenwood, R. C., Franchi, I. A., Jambon, A., Barrat, J. A. & Burbine, T. H. Oxygen isotope variation in stony-iron meteorites. *Science* **313**, 1763–1765 (2006).

6. Trinquier, A., Birck, J.-L. & Allègre, C. G. Widespread $^{54}$Cr heterogeneities in the inner solar system. *Astrophys. J.* **655**, 1179–1185 (2007).

7. Trinquier, A., Birck, J.-L., Allegre, C.-J., Göpel, C. & Ulfbeck, D. $^{53}$Mn–$^{53}$Cr systematics of the early Solar System revisited. *Geochim. Cosmochim. Acta* **72**, 5146–5163 (2008).

8. McCord, T. B., Adams, J. B. & Johnson, T. V. Asteroid Vesta: spectral reflectivity and compositional implications. *Science* **168**, 1445–1447 (1970).

9. De Sanctis, M. C. *et al.* Spectroscopic characterization of mineralogy and its diversity across vesta. *Science* **336**, 697–700 (2012).

10. Prettyman, T. H. *et al.* Elemental mapping by Dawn reveals exogenic H in Vesta's regolith. *Science* **338**, 242–246 (2012).

11. Rubin, A. E. & Mittlefehldt, D. W. Evolutionary history of the mesosiderite asteroid: a chronologic and petrologic synthesis. *Icarus* **101**, 201−212 (1993).

12. Stewart, B. W., Papanastassiou, D. A. & Wasserburg, G. J. Sm-Nd chronology and petrogenesis of mesosiderites. *Geochim. Cosmochim. Acta* **58**, 3487–3509 (1994).

13. Ireland, T. R. & Wlotzka, F. The oldest zircons in the solar system. *Earth Planet. Sci. Lett*. **109**, 1–10 (1992).

14. Haba, M. K., Yamaguchi, A., Kagi, H. Nagao, K. & Hidaka, H. Trace element composition and U-Pb age of zircons from Estherville: Constraints on the timing of the metal-silicate mixing event on the mesosiderite parent body. *Geochim. Cosmochim. Acta* **215,** 76–91 (2017).

15. Koike, M. Sugiura, N. Takahata, N. Ishida, A. & Sano, Y. U-Pb and Hf-W dating of young zircon in mesosiderite Asuka 882023. *Geophys. Res. Lett*., **44**, 1251–1259 (2017).

16. Vermeesch, P. Dissimilarity measures in detrital geochronology. *Earth-Sci. Rev*. **178**, 310–321 (2018).

17. Hewins, R. H. The case for a melt matrix in plagioclase-POIK mesosiderites. *J. Geophys. Res.* **89** (suppl.), C289–C297 (1984).

18. Delaney, J. S., Nehru, C. E., Prinz, M. & Harlow, G. E. Metamorphism in mesosiderites. *Proc. Lunar Planet. Sci. Conf.* **12,** 1315–1342 (1981).

19. Prinz, M., Nehru, C. E., Delaney, J. S., Harlow, G. E. & Bedell, R. L. Modal studies of mesosiderites and related achondrites, including the new mesosiderite ALHA 77219. *Proc. Lunar Planet. Sci. Conf. 11th*, 1055–1071 (1980).

20. Delaney, J. S., Prinz, M. & Takeda, H. The polymict eucrites. *Proc. Lunar Planet. Sci. Conf. 15th, in J. Geophys. Res.,* **89**, C251–C288 (1984).

21. Harlow, G. E., Delaney, J. S., Nehru, C. E. & Prinz, M. Metamorphic reactions in mesosiderites: origin of abundant phosphate and silica. *Geochim. Cosmochim. Acta* **46**, 339–348 (1982).

22. Rubin, A. E. & Mittlefehldt, D. W. Classification of mafic clasts from mesosiderites: implications for endogenous igneous processes. *Geochim. Cosmochim. Acta* **56**, 827–840 (1992).

23. Misawa, K., Yamaguchi, A. & Kaiden, H. U-Pb and $^{207}$Pb-$^{206}$Pb ages of zircons from basaltic eucrites: implications for early basaltic volcanism on the eucrite parent body. *Geochim. Cosmochim. Acta* **69**, 5847–5861 (2005).

24. Zhou, Q. *et al*. SIMS Pb-Pb and U-Pb age determination of eucrite zircons at <5 mm scale and the first 50 Ma of the thermal history of Vesta. *Geochim. Cosmochim. Acta* **110**, 152–175 (2013).

25. Iizuka, T. *et al*. Timing of global crustal metamorphism on Vesta as revealed by high-precision U-Pb dating and trace element chemistry of eucrite zircon. *Earth Planet. Sci. Lett.* **409**, 182–192 (2015).

26. Hopkins, M., Mojzsis, S., Bottke, W. & Abramov, O. Micrometer-scale U-Pb age domains in eucrite zircons, impact re-setting, and the thermal history of the HED parent body. *Icarus* **245**, 367–378 (2015).

27. Srinivasan, G., Whitehouse, M. J., Weber, I. & Yamaguchi, A. The crystallization age of eucrite zircon. *Science* **317**, 345–347 (2007).

28. Ireland, T. R., Saiki, K. & Takeda, H. Age and trace-element chemistry of Yamato-791438 zircon. *Lunar Planet. Sci.* abstr. **23,** 569–570 (1992).

29. Roszjar, J. *et al*. Prolonged magmatism on 4 Vesta inferred from Hf-W analyses of eucrite zircon. *Earth Planet. Sci. Lett.* **452**, 216–226 (2016).

30. Roszjar, J. *et al*. Thermal history of Northwest Africa (NWA) 5073 – a coarse grained Stannern-trend eucrite containing cm-sized pyroxenes and large zircon grains. *Meteor. Planet. Sci.* **46**, 1754–1773 (2011).

31. Shukolyukov, A. & Begemann, F. Pu-Xe dating of eucrites. *Geochim. Cosmochim. Acta* **60**, 2453–2480 (1996).

32. Premo, W. R. & Tatsumoto, M. U-Th-Pb, Rb-Sr, and Sm-Nd isotopic systematics of lunar troctolitic cumulate 76535: Implications on the age and origin of this early lunar, deep-seated cumulate. *Proc. 22nd Lunar Planet. Sci. Conf.* 38l–397 (1992).

33. Haba, M. K., Yamaguchi, A., Horie, K. & Hidaka, H. Major and trace elements of zircons from basaltic eucrites: implications for the formation of zircons on the eucrite parent body. *Earth Planet. Sci. Lett.* **387**, 10–21 (2014).

34. Carter, P. J., Leinhardt, Z. M., Elliott, T., Stewart, S. T. & Walter, M. J. Collisional stripping of planetary crusts. *Earth Planet. Sci. Lett*. **484**, 276–286 (2018).

35. Asphaug, E. Similar-sized collisions and the diversity of planets. *Chem. Erde Geochem.* **70**, 199–219 (2010).

36. Ammannito, E. *et al.* Olivine in an unexpected location on Vesta's surface. *Nature* **504**, 122–125 (2013).

37. Ruesch, O. *et al.* Detections and geologic context of local enrichments in olivine on Vesta with VIR/Dawn data, *J. Geophys. Res. Planets, 119*, 2078–2108 (2014).

38. Clenet, H. *et al.* A deep crust-mantle boundary in the asteroid 4 Vesta. *Nature* **511**, 303–306 (2014).

39. Ruzicka, A., Snyder, G. A. & Taylor, L. Vesta as the HED parent body: implications for the size of a core and for large-scale differentiation. *Meteorit. Planet. Sci.* **32**, 825−840 (1997).

40. Mandler, B. E. & Elkins-Tanton, L. T. The origin of eucrites, diogenites, and olivine diogenites: magmaocean crystallization and shallow magma chamber processes on Vesta. *Meteorit. Planet. Sci.* **48**, 1–17 (2013).

41. Barrat, J.-A., Yamaguchi, A., Zanda, B., Bollinger, C. & Bohn, M. Relative chronology of crust formation on asteroid Vesta: insights from the geochemistry of diogenites. *Geochim. Cosmochim. Acta* **74**, 6218−6231 (2010).

42. Consolmagno, G. J. *et al.* Is Vesta an intact and pristine protoplanet? *Icarus* **254**, 190–201 (2015).

43. Hopfe, W. D. & Goldstein, J. I. The metallographic cooling rate method revised: Application to iron meteorites and mesosiderites. *Meteorit. Planet. Sci*. **36**, 135–154 (2001).

44. Bogard, D. D. & Garrison, D. H. $^{39}$Ar-$^{40}$Ar ages and thermal history of mesosiderites. *Geochim. Cosmochim. Acta* **62**, 1459–1468 (1990).

45. Schenk, P. *et al*. The geologically recent giant impact basins at vesta's south pole. *Science* **336**, 694–697 (2012).

46. Jutzi, M., Asphang, E., Gillet, P., Barrat, J. -A. & Benz, W. The structure of the asteroid 4 Vesta as revealed by models of planet-scale collisions. *Nature* **494**, 207–210 (2013).

47. Binzel, R. P. & Xu, S. Chips off of asteroid 4 Vesta: evidence for the parent body of basaltic achondrite meteorites. *Science* **260**, 186–191 (1993).

48. Nesvorný, D. *et al*. Fugitives from the Vesta family. *Icarus* **193**, 85–95 (2008).

49. Condon, D. J., Schoene, B., McLean, N. M., Bowring, S. A. & Parrish, R. R. Metrology and traceability of U-Pb isotope dilution geochronology (EARTHTIME Tracer Calibration Part I). *Geochim. Cosmochim. Acta* **164**, 464–480 (2015).

50. Krogh, T. E. A low-contamination method for hydrothermal decomposition of zircon and extraction of U and Pb for isotopic age determinations. *Geochim. Cosmochim. Acta* **37**, 488–494 (1973).

51. Gerstenberger, H. & Haase, G. A highly effective emitter substance for mass spectrometric Pb isotope ratio determinations. *Chem. Geol.* **136**, 309–312 (1997).

52. von Quadt, A. *et al.* High-precision zircon U/Pb geochronology by ID-TIMS using new $10^{13}$ ohm resistors. *J. Anal. At. Spectrom.* **31**, 658–665 (2016).

53. Wotzlaw, J. F., Buret, Y., Large, S. J. E., Szymanowski, D. & von Quadt, A., ID-TIMS U–Pb geochronology at the 0.1‰ level using $10^{13}$ Ω resistors and simultaneous U and $^{18}O/^{16}O$ isotope ratio determination for accurate $UO_2$ interference correction. *J. Anal. At. Spectrom.* **32**, 579–586 (2017).

54. Bowring, J. F., McLean, N. M. & Bowring, S. A. Engineering cyber infrastructure for U-Pb geochronology: Tripoli and U-Pb_Redux. *Geochem. Geophys. Geosyst.* **12**, Q0AA19 (2011).

55. McLean, N. M., Bowring, J. F. & Bowring, S. A. An algorithm for U-Pb isotope dilution data reduction and uncertainty propagation. *Geochem. Geophys. Geosyst.* **12**, Q0AA18 (2011).

56. Jaffey, A. H., Flynn, K. F., Glendenin, L. E., Bentley, W. C. & Essling, A. M. Precision measurement of half-lives and specific activities of $^{235}U$ and $^{238}U$. *Phys. Rev. C* **4**, 1889–1906 (1971).

57. Connelly, J. N. *et al.* The Absolute Chronology and Thermal Processing of Solids in the Solar Protoplanetary Disk. *Science* **338**, 651–655 (2012).

58. Brennecka, G. A. & Wadhwa, M. Uranium isotope compositions of the basaltic angrite meteorites and the chronological implications for the early Solar System. *Proc. Natl Acad. Sci. USA* **109**, 9299–9303 (2012).

**Acknowledgments**

The authors thank H. Genda for insightful discussions. M.K.H. acknowledges support from JSPS Postdoctoral Fellowship for Research Abroad (No. 27-699), J.F.W. from ETH Zurich postdoctoral fellowship program (FEL-14-09), Y.-J.L., and M.S. from the European Research Council under the European Union's Seventh Framework Programme (FP7/2007–2013)/ERC Grant agreement No. [279779] and the Swiss National Science Foundation (Project 200021_149282), and A.Y. from NIPR Research Project KP307.

**Author contributions**

M.K.H. and M.S. designed the research. M.K.H., Y.-J. L., and A.Y. prepared zircon samples. J.F.W. performed U-Pb dating of zircons. M.K.H. took the lead in writing the manuscript. All authors discussed the results and commented on the manuscript.

**Competing interests**

The authors declare no competing interests.

**Data availability**

The data that support the findings of this study are available from the corresponding author upon reasonable request.

**Figures**

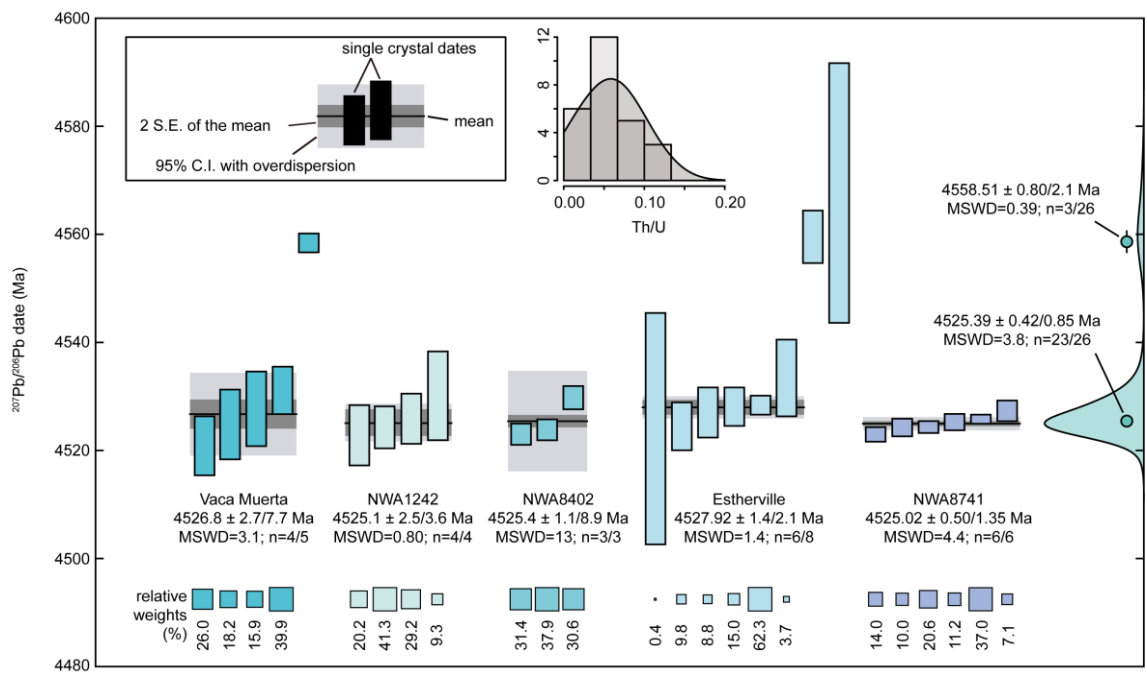

**Figure 1. $^{207}$Pb/$^{206}$Pb dates for mesosiderite zircons.** Each panel shows ranked $^{207}$Pb/$^{206}$Pb dates with their 2σ uncertainties. Given are the weighted mean dates for each mesosiderite and corresponding mean square weighted deviation (MSWD). Uncertainties are given as 2 S.E. and as studentised 95% confidence intervals including overdispersion[16]. Lower panel illustrates the relative weights of individual dates that were included in the weighted mean calculation. Shown on the right is a Kernel density estimate of the entire data set with the weighted mean of the two resolved zircon populations. Inset depicts the distribution of Th/U ratios of analysed zircons. All uncertainties are given at the 95% confidence level and do not include decay constant uncertainties.

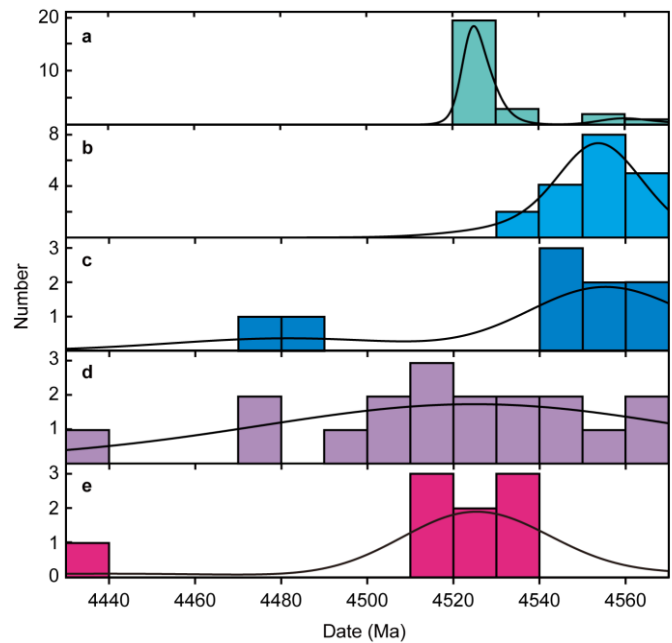

**Figure 2. Histograms and Kernel density estimates of mesosiderite zircon dates and various dates of basaltic eucrite. a,** $^{207}$Pb-$^{206}$Pb dates mesosiderite zircons (see Fig. 1 and Supplementary Table 1 for details). **b,** $^{206}$Pb-$^{207}$Pb and $^{182}$Hf-$^{182}$W dates of zircons from basaltic eucrites (N=17). **c,** $^{147,146}$Sm-$^{142,143}$Nd mineral isochron dates of nine basaltic eucrites. **d,** $^{244}$Pu-Xe whole rocks dates of 17 basaltic eucrites. **e,** $^{207}$Pb-$^{206}$Pb whole rock dates or mineral isochron dates of 8 basaltic eucrites. Black solid lines show Kernel density estimates. All data including references used for basaltic eucrites are summarized in Supplementary Table 2.

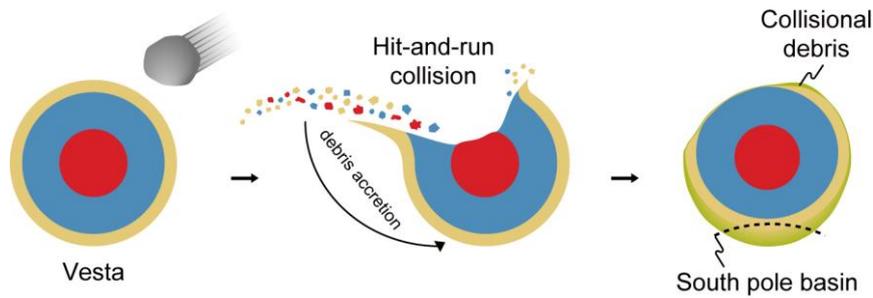

**Figure 3. Mesosiderite formation on Vesta in an internal-origin model adopting a hit-and-run collision.** This model assumes a hit-and-run collision of Vesta as the parent body of mesosiderite metal and silicate with a smaller planetesimal (mass ratio 0.1). This collision caused large-scale disruption of the northern hemisphere and ejected debris composed of crust (yellow), mantle (blue), and core materials (red). The debris accreted to the opposite surface of the collision site. Thickened layer, composed of original crust and mixture of collisional debris (green), overlying mantle in the southern hemisphere corresponding to the current Veneneia and Rheasilvia basins. The approximate location of the south pole basin is shown as dashed line.

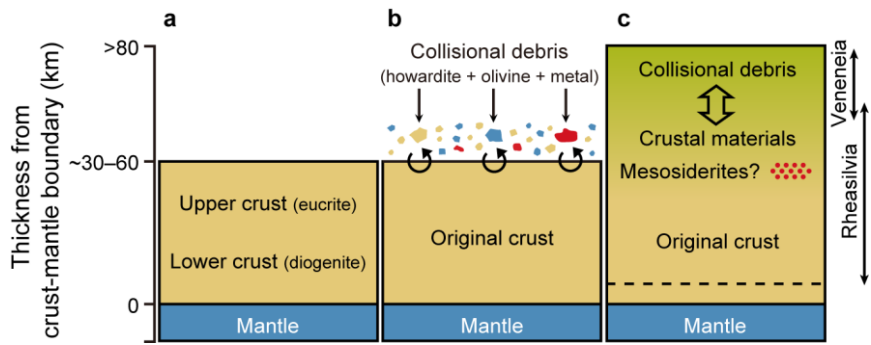

**Figure 4. Crust evolution in the south pole region. a**, Original crust composed of eucritic upper crust and diogenitic lower crust. **b**, Accretion of hit-and-run collisional debris on the surface. The debris was fragmentated into smaller pieces and underwent mixing with the underlaying crust in the early stage. **c**, Thickened layer on mantle after the collision. Mesosiderite formation area is shown as red dots. Two giant impacts that formed the Veneneia and Rheasilvia basins, ejected materials from upper layers enriched in collisional debris and lower ones dominated by the original crust, respectively. The dashed line indicates the lower limit of ejection. Each panel corresponds to the collision process in Fig. 3 (same colour code).

**Supplementary Information**

**Additional Results**

<u>Occurrence, internal texture, and crystallinity of mesosiderite zircons</u>

Polished thick sections (10×10×1 mm) of mesosiderites (Vaca Muerta, NWA 1242, NWA 8402, Estherville, NWA 8741) were prepared from the sample samples used for zircon separation for ID-TIMS U-Pb dating. Zircons were identified by elemental mapping of Zr Lα using an electron probe micro analyzer (JEOL JXA 8200) at the National Institute of Polar Research (NIPR). The representative elemental mapping of a mesosiderite sample is shown in Fig. S1. Back-scattered electron (BSE) and cathodoluminescence (CL) images of the zircons were obtained using a field emission scanning electron microscope (FE-SEM, JEOL JSM-7100F) with a CL detector at NIPR. A few zircons were found in each thick section except for NWA 1242 (Fig. S2). All the zircons are located in the silicate parts or at the boundary between silicate and Fe-Ni metal parts, contacting with Fe-Ni metal or troilite (Figs. S1 and S2).

A few remaining zircon grains (>40 μm) were mounted in epoxy resin and polished with 3 μm and 1/4 μm diamond pastes. All the zircons from NWA 8402 were consumed for the ID-TIMS U-Pb dating. Internal texture of zircon was observed using the FE-SEM-CL system. Whereas the zircons in basaltic eucrites are mostly less than 20–30 μm (ref. 33), the mesosiderites contain the zircons up to ~180 μm (Supplementary Table 1). Previous work documented a zircon in the Estherville mesosiderite with heterogeneous internal texture in BSE and CL images, corresponding to different concentrations of U, Th, and rare earth elements[14]. All the zircons observed in this study show homogeneous internal texture in both of the BSE and CL images (Fig. S3). This result indicates that these zircons have relatively homogeneous trace element concentrations.

Raman spectra of the zircons used for the ID-TIMS U-Pb dating and remaining zircons were obtained with a Raman microscope (Renishaw inVia) with excitation at 532 nm. Raman shift was calibrated using the Si Raman peak at 520 $cm^{-1}$. The zircons for the Raman analysis were limited to those having grain size larger than 40 μm because the smaller grains are difficult to transfer to the sample container without any loss. Representative Raman spectra of the mesosiderite zircons are shown in Fig. S4 together with that of terrestrial zircon 91500. The highest peak of the terrestrial zircon 91500 was observed at 1005.6 cm$^{-1}$ with the full width at half maximum (FWHM) of 5.26 cm$^{-1}$. Mesosiderite zircons show the highest peak at slightly higher frequencies (1007.2–1007.4 cm$^{-1}$) and narrower FWHM (4.48–4.57 cm$^{-1}$) compared with those of the terrestrial zircon 91500. All the measured zircons (>40 μm) have similar Raman spectra shown in Fig. S4. It has been known that the peak around 1000 cm$^{-1}$ ($v_3$(SiO$_4$) band) shifts to lower wavenumber and becomes broader due to α-decay damage from U and Th over geological time[59]. The high frequency and narrow FWHM of $v_3$(SiO$_4$) band of the mesosiderite zircons indicate that they experienced less radiation damage than the terrestrial zircon 91500. The α dose of terrestrial zircon 91500 is 3.3×10$^{17}$ α-decay/g based on the U and Th concentrations (80.0 and 29.9 ppm) (ref. 60) and formation age (1,065 Myr) (ref. 61). The comparably low α-dosages of mesosiderite zircons is consistent with the low U content, suggesting that radiation damage is limited and corresponding Pb-loss from radiation damaged domains[62] is unlikely.

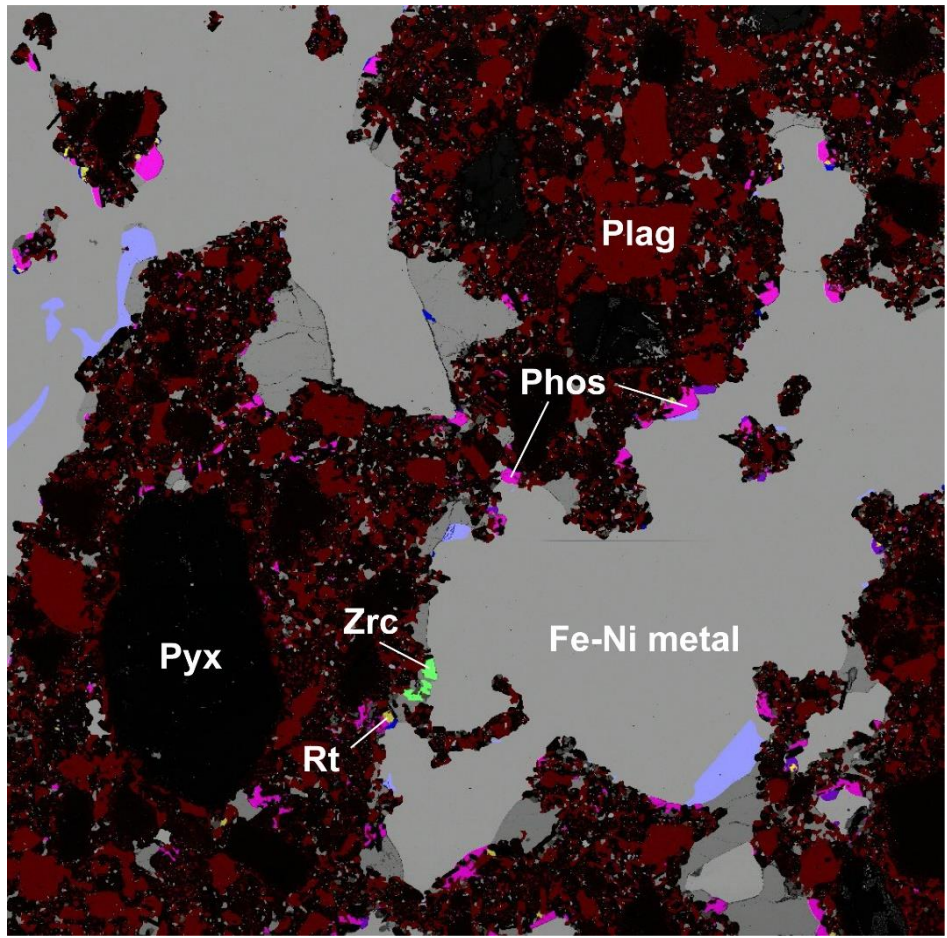

**Figure S1. Elemental mapping on the back-scattered electron (BSE) image of the Estherville mesosiderite.** Zircon, rutile, and phosphate mineral are shown as green, yellow, pink coloured areas, respectively (Zr: green, Ti: yellow, Ca: red, P: blue). Most of these minerals are located at the boundary between the silicate and metal. This feature can be observed in all types of mesosiderites. The width of photos is 1 cm. Mineral abbreviations are as follows: Zrc, zircon; Rt, rutile; Phos, phosphate mineral; Pyx, pyroxene; Plag, plagioclase.

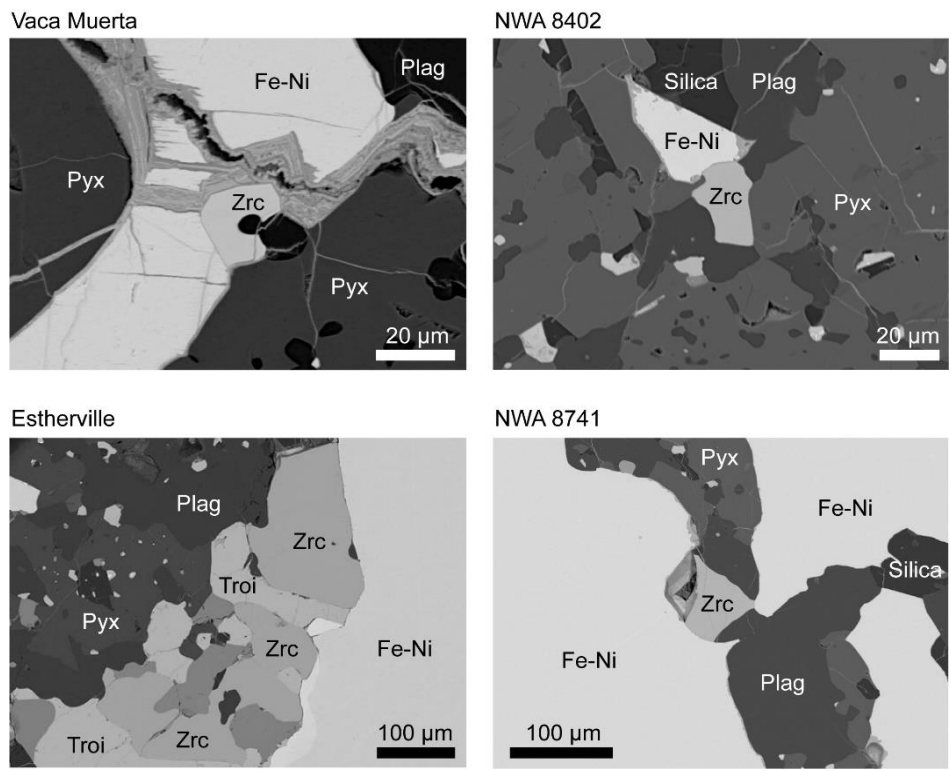

**Figure S2. BSE images of mesosiderites (Vaca Muerta, NWA 8402, Estherville, NWA 8741) used in this study.** Mineral abbreviations are as follows: Zrc, zircon; Troi, troilite; Pyx, pyroxene; Plag, plagioclase.; Silica, silica mineral.

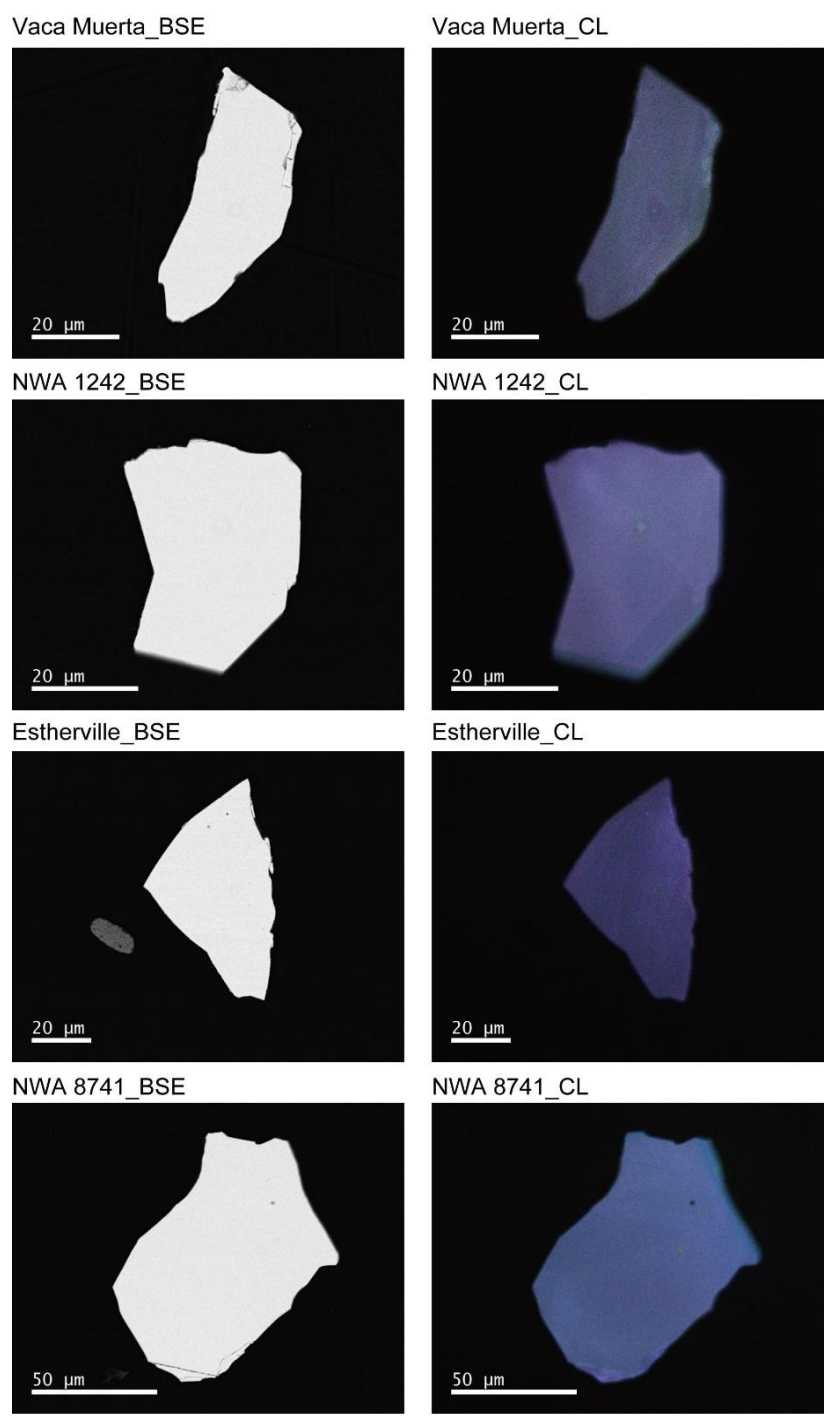

**Figure S3. BSE and CL images of mesosiderites zircons (Vaca Muerta, NWA 1242, Estherville, NWA 8741) used in this study.** The small circles (~2 μm) in CL images are beam irradiation marks from the EPMA measurement.

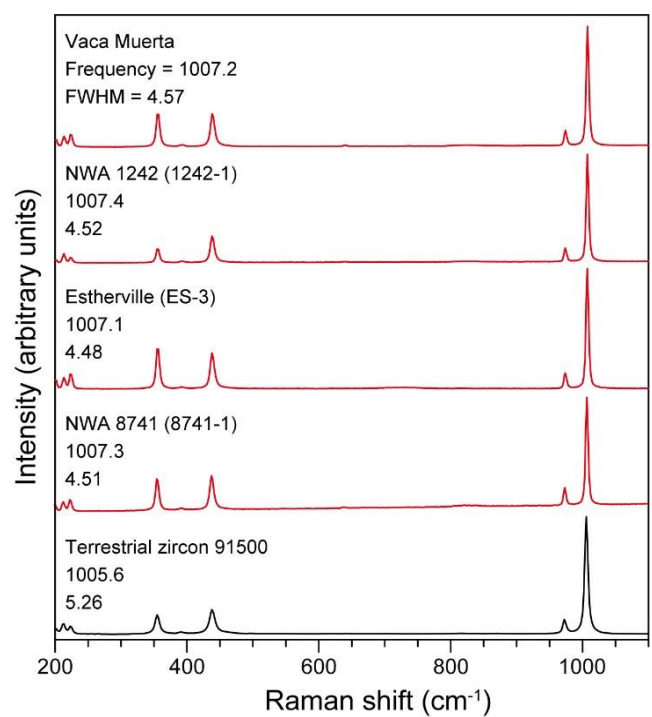

**Figure S4. Representative Raman spectra between 200 and 1100 cm$^{-1}$ of zircons from mesosiderites and terrestrial zircon 91500.** Frequency and full width at half maximum (FWHM) of $\nu_3(SiO_4)$ band of each zircon are shown in the figure.

**Supplementary Discussion**

Highly depleted incompatible elements and high Eu/Sm ratios in gabbroic clasts

Mesosiderite silicates have experienced thermal metamorphism at equal to or greater than solidus temperature (~1100°C) (ref. 63) at the metal-silicate mixing. This process caused partial melting of silicates and redox reaction between silicate and molten metal, and thereby abundant secondary phosphate minerals formed in the silicate parts[21]. In addition, ilmenite which is a host mineral of Ti, Zr, Hf, Nb, and Ta also dissolved during the thermal metamorphism because of its lower melting point (1050–1100°C) (ref. 64) and resulted in forming secondary zircon and rutile[14]. As reported in ref. 64, the minerals having low melting point preferentially dissolve and redistribute in the samples during thermal metamorphism around solidus temperature. Indeed, phosphate minerals, zircon, and rutile are mostly found at the boundary of silicate and metal parts in mesosiderites (Fig. S1). Therefore, in order to determine the original chemical compositions of mesosiderite silicate clasts, the sample must include the minerals at the boundary between silicate and metal parts. The gabbroic clasts in the Vaca Muerta mesosiderite, whose metal contents are equal or less than 0.5 wt.%, have shown highly depleted incompatible elements and high Eu/Sm ratio. Such incompatible element compositions have never been observed in HED meteorites. In Fig. S5, we have reproduced the REE compositions of mesosiderite clasts having highly fractionated Eu/Sm ratio (pebble 12, 17, 18) in ref. 22 by assuming a cumulate eucrite as an original silicate, considering the effect of unsampled phosphate minerals. As a result, such a highly fractionated Eu/Sm ratio and low incompatible element concentrations can be explained by the effect of unsampled phosphate minerals (for REE) and possibly rutile and zircon (for Ti, Zr, Nb, Hf, Ta). Thus, incompatible element compositions of the gabbroic clasts are likely related with heterogeneous distribution of the secondary minerals formed during metal-silicate mixing and do not deny the genetic linkage of mesosiderites and HED meteorites. The secondary phosphate mineral shown in Fig. S1 are considered to have formed in contact with or very close to molten metal to get phosphorus. Since HED meteorites contain only trace amount of metal, the redox reaction related with the molten metal would have not occurred in the area where HED resided on the parent body.

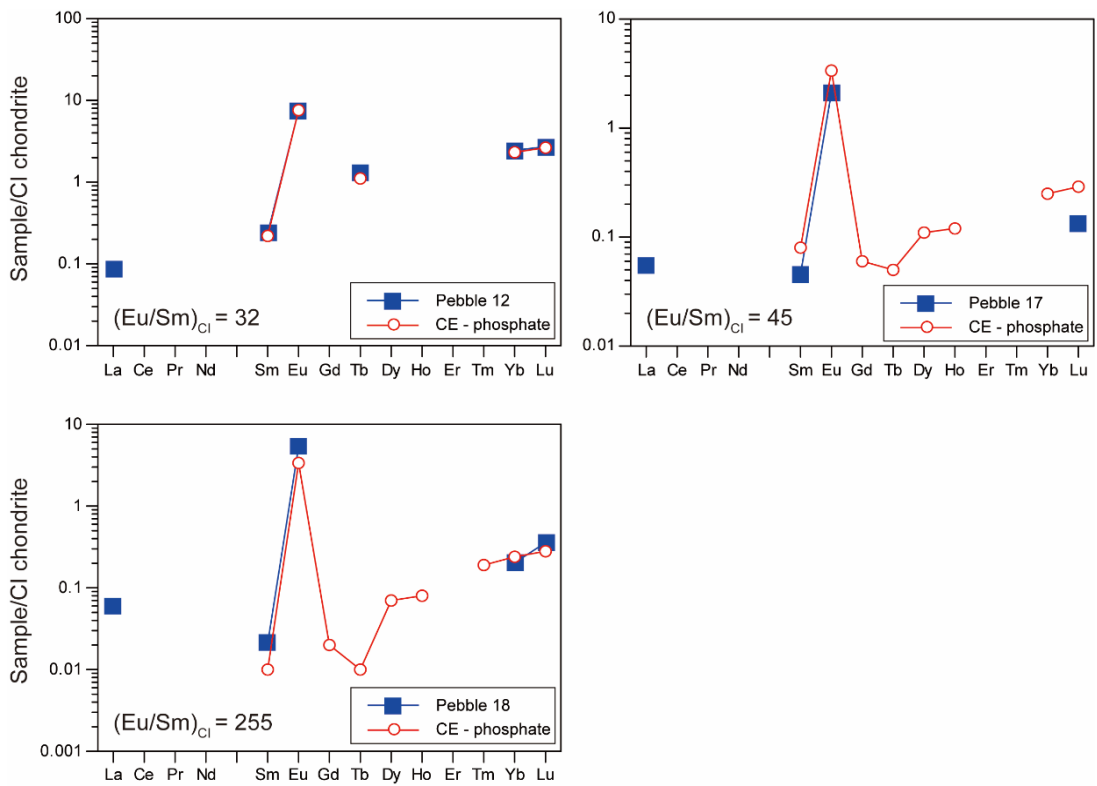

**Figure S5. Reproduction of the REE concentrations of pebbles 12, 17, and 18 in the Vaca Muerta mesosiderite reported in ref. 22.** In the calculation, we have assumed that the original silicate is a cumulate eucrite (CE) and used the REE data of Asuka 881819 (ref. 65) for Pebble 12 and Medanitos[66] for Pebbles 17 and 18. The REE component of unsampled phosphate minerals was subtracted from the REE of cumulate eucrites so as to fit to the reported REE data of each pebble. The REE concentrations of merrillite from Pomozdino[67] were used in the calculation. The calculated data can mostly reproduce the highly fractionated Eu/Sm ratio and REE concentrations of the three pebbles, except for La. In the calculation, La concentrations become much lower than the reported values. Since light REE concentrations of meteorite finds can be elevated by terrestrial weathering[68], the La concentrations observed in pebbles 12, 17, and 18 of Vaca Muerta (meteorite find) are likely to have been affected by terrestrial weathering.

Parent body size of mesosiderite metal

The $^{207}$Pb-$^{206}$Pb age of younger zircons dating metal-silicate mixing entails that the core material that formed mesosiderite metal was still almost completely molten 4,525.4 Myr ago. According to the crystallization ages of iron meteorites (IIIAB, IIAB, and IVB) dated by the $^{187}$Re-$^{187}$Os chronometer, the cores of the parent bodies of these meteorites, whose radii are less than 100 km, already crystallized ~4,520 Myr ago[69,70]. Similarly for the IAB meteorite parent body, it has been shown that bodies smaller than 60 km crystallized a solid metal core 4,557 Myr ago[71]. Numerical calculations of thermo-chemical evolution for Vesta (~530 km in diameter) imply that 60–80% of Vesta's core was molten around 4,525.4 Myr ago if the accretion of Vesta started within 1 Myr after CAI formation[72]. Even after the consideration of uncertainties in the numerical calculation, the parent body of mesosiderite metal is estimated to be equal or larger than the current Vesta. This entails that the parent body of the mesosiderite metal was a protoplanet ≥530 km in diameter and experienced a catastrophic collision and break-up 4,525.39 ± 0.85 Myr ago.

**Supplementary Figure**

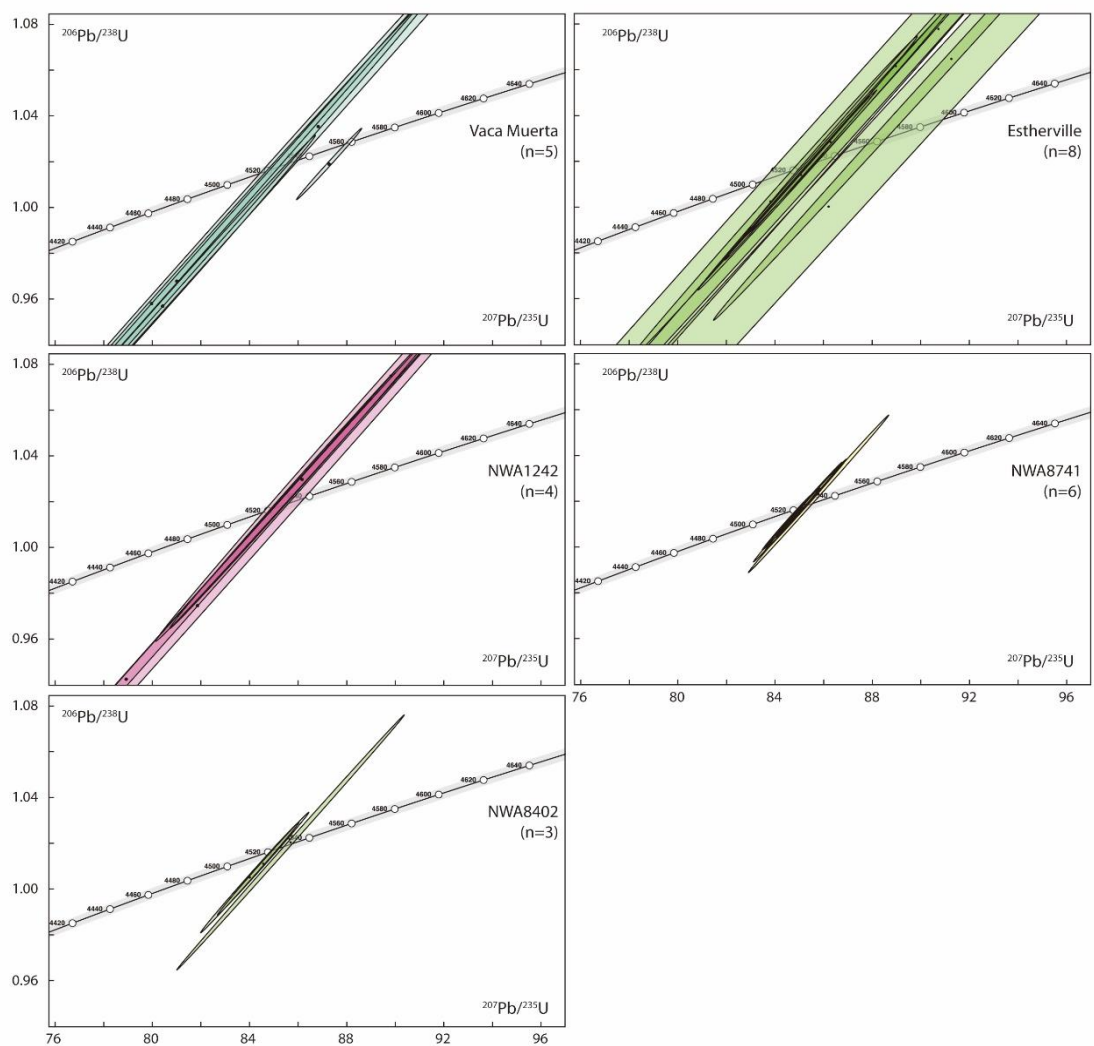

**Figure S6. U-Pb concordia diagrams for mesosiderite zircons (Vaca Muerta, NWA 1242, NWA 8402, Estherville, NWA 8741).** Error ellipses are 2σ confidence level.

# Supplementary Table

**Table S1a.** U-Pb isotopic data of zircons and aliquots from Vaca Muerta and NWA 1242 mesosiderites.

| Sample | Diameter (μm) | Weight [a] (μg) | Th/U [b] | U [c] (pg) | U [d] (ppm) | Pb* [e] (pg) | Pb$_c$ [f] (pg) | Pb*/Pb$_c$ [g] | $^{206}$Pb/$^{204}$Pb [h] | $^{208}$Pb/$^{206}$Pb [i] | $^{206}$Pb/$^{238}$U [i] | ± 2σ (%) | $^{207}$Pb/$^{235}$U [i] | ± 2σ (%) | $^{207}$Pb/$^{206}$Pb [i] | ± 2σ (%) |
|---|---|---|---|---|---|---|---|---|---|---|---|---|---|---|---|---|
| ET2Ga (2 Ga synthetic solution) | | | | | | | | | | | | | | | | |
| 16594 | | | 0.091 | 144 | | 51.6 | 0.326 | 158.25 | 9917.8 | 0.026 | 0.36376 | 0.045 | 6.1686 | 0.086 | 0.123045 | 0.06 |
| 16650 | | | 0.091 | 143 | | 51.4 | 0.317 | 161.92 | 10147.2 | 0.026 | 0.36362 | 0.042 | 6.1682 | 0.079 | 0.123084 | 0.05 |
| 17312 | | | 0.092 | 142 | | 51.0 | 0.382 | 133.64 | 8378.1 | 0.026 | 0.36352 | 0.042 | 6.1644 | 0.076 | 0.123044 | 0.05 |
| 17313 | | | 0.091 | 134 | | 48.1 | 0.357 | 134.71 | 8446.0 | 0.026 | 0.36362 | 0.046 | 6.1650 | 0.078 | 0.123021 | 0.04 |
| 17314 | | | 0.092 | 234 | | 84.3 | 0.535 | 157.57 | 9875.3 | 0.026 | 0.36352 | 0.037 | 6.1640 | 0.065 | 0.123033 | 0.04 |
| Vaca Muerta (type 1) | | | | | | | | | | | | | | | | |
| VM-1 | ~40 | | 0.064 | 0.7 | | 0.37 | 0.083 | 4.42 | 217.5 | 0.018 | 0.96 | 16.20 | 80.0 | 16.2 | 0.6057 | 0.37 |
| VM-2 | ~40 | | 0.015 | 0.1 | | 0.18 | 0.054 | 3.31 | 165.9 | 0.004 | 1.04 | 18.09 | 86.8 | 18.1 | 0.6087 | 0.47 |
| VM-3 | ~40 | | 0.081 | 0.3 | | 0.41 | 0.081 | 5.04 | 240.3 | 0.020 | 0.96 | 7.81 | 80.4 | 7.8 | 0.6100 | 0.30 |
| VM-4 | ~40 | | 0.081 | 0.3 | | 0.31 | 0.085 | 3.68 | 180.4 | 0.021 | 0.97 | 13.21 | 81.0 | 13.2 | 0.6074 | 0.44 |
| VM-5 (k) | ~40, 40 | | 0.100 | 1.9 | | 2.06 | 0.078 | 26.3 | 1167.8 | 0.026 | 1.02 | 1.53 | 87.3 | 1.5 | 0.62162 | 0.11 |
| NWA 1242 (type 2) | | | | | | | | | | | | | | | | |
| 1242-1 | 50–60 | | 0.020 | 0.4 | | 0.52 | 0.082 | 6.32 | 300.8 | 0.005 | 1.03 | 6.33 | 86.2 | 6.3 | 0.6074 | 0.27 |
| 1242-2 | 50–60 | | 0.009 | 0.2 | | 0.32 | 0.084 | 3.79 | 187.0 | 0.002 | 1.07 | 10.77 | 89.8 | 10.8 | 0.6065 | 0.38 |
| 1242-3 | 50–60 | | 0.035 | 2.6 | | 0.40 | 0.080 | 5.05 | 241.9 | 0.011 | 0.94 | 26.82 | 78.9 | 26.8 | 0.6079 | 0.32 |
| 1242-4 | 50–60 | | 0.014 | 0.3 | | 0.39 | 0.153 | 2.52 | 130.4 | 0.004 | 0.97 | 20.89 | 81.9 | 20.9 | 0.6096 | 0.56 |

**Table S1a.** (*Continued*)

| Sample | Dates (Ma) | | | | | | MSWD |
|---|---|---|---|---|---|---|---|
| | $^{206}Pb/^{238}Pb$ [j] | ± 2σ (abs) | $^{207}Pb/^{235}Pb$ [j] | ± 2σ (abs) | $^{207}Pb/^{206}Pb$ [j] | ± 2σ (abs) | |
| ET2Ga (2 Ga synthetic solution) | | | | | | | |
| 16594 | 1999.99 | 0.77 | 2000.02 | 0.75 | 2000.0 | 1.1 | |
| 16650 | 1999.31 | 0.72 | 1999.95 | 0.69 | 2000.6 | 1.1 | |
| 17312 | 1998.82 | 0.72 | 1999.41 | 0.66 | 2000.0 | 1.0 | |
| 17313 | 1999.30 | 0.79 | 1999.50 | 0.68 | 1999.70 | 0.98 | |
| 17314 | 1998.85 | 0.63 | 1999.36 | 0.56 | 1999.88 | 0.89 | |
| Vaca Muerta (type 1) | | | | | | | |
| VM-1 | 4332 | 511 | 4462 | 163 | 4520.8 | 5.4 | |
| VM-2 | 4581 | 593 | 4544 | 182 | 4527.8 | 6.9 | |
| VM-3 | 4328 | 246 | 4467 | 78 | 4531.1 | 4.3 | |
| VM-4 | 4364 | 419 | 4475 | 133 | 4524.8 | 6.4 | |
| VM-5 [k] | 4529 | 50 | 4549 | 15 | 4558.3 | 1.7 | |
| | | | **Weighted mean withoout VM-5** | | **4526.8** | **2.7** | **3.1** |
| NWA 1242 (type 2) | | | | | | | |
| 1242-1 | 4563 | 207 | 4537 | 64 | 4524.7 | 3.9 | |
| 1242-2 | 4705 | 360 | 4578 | 108 | 4522.7 | 5.6 | |
| 1242-3 | 4280 | 839 | 4449 | 269 | 4525.9 | 4.6 | |
| 1242-4 | 4386 | 665 | 4485 | 210 | 4530.1 | 8.2 | |
| | | | **Weighted mean** | | **4525.1** | **2.5** | **0.8** |

[a] Weight of zircon estimated from crystal size by assuming spherical shape and density of 4.7 g/cm$^3$.

[b] Th content calculated from radiogenic $^{208}$Pb and $^{207}$Pb/$^{206}$Pb date of the sample, assuming concordance between U-Pb Th-Pb systems.

[c] U amount measured by ID-TIMS.

[d] U concentration calculated from the U amount measured by ID-TIMS (c) and the zircon weight (a).

[e] Total mass of radiogenic Pb (*).

[f] Total mass of common Pb.

[g] Ratio of radiogenic Pb (including $^{208}$Pb) to common Pb.

[h] Measured ratio corrected for fractionation and spike contribution only.

[i] Measured ratios corrected for fractionation, tracer, and blank.

[j] Isotopic dates calculated using $\lambda_{238} = 1.55125 \times 10^{-10}$ and $\lambda_{235} = 9.8485 \times 10^{-10}$ (ref. 56, see Methods).

[k] Aliquots which contain two or three zircons. ES-6 contains three grains including two small grains (<40 μm).

MSWD: mean square weighted deviation.

Grain sizes of zircons from NWA 8402 and NWA 8741were determined using video measuring system attached to a stereoscopic microscope.

**Table S1b.** U-Pb isotopic data of zircons and aliquots from NWA 8402, Estherville, and NWA 8741 mesosiderites.

| Sample | Diameter (μm) | Weight [a] (μg) | Th/U [b] | U [c] (pg) | U [d] (ppm) | Pb* [e] (pg) | Pb$_c$ [f] (pg) | Pb*/Pb$_c$ [g] | Isotopic ratios $^{206}$Pb/$^{204}$Pb [h] | $^{208}$Pb/$^{206}$Pb [i] | $^{206}$Pb/$^{238}$U [i] | ± 2σ (%) | $^{207}$Pb/$^{235}$U [i] | ± 2σ | $^{207}$Pb/$^{206}$Pb [i] | ± 2σ (%) |
|---|---|---|---|---|---|---|---|---|---|---|---|---|---|---|---|---|
| NWA 8402 (type 3) | | | | | | | | | | | | | | | | |
| 8402-1 | 69 × 129 | 1.5 | 0.002 | 0.9 | 0.6 | 1.27 | 0.09 | 14.6 | 667.7 | 0.001 | 1.01 | 2.41 | 84.0 | 2.4 | 0.6067 | 0.14 |
| 8402-2 | 88 × 100 | 1.9 | 0.036 | 1.2 | 0.6 | 1.59 | 0.10 | 16.5 | 747.5 | 0.009 | 1.01 | 2.22 | 84.6 | 2.2 | 0.6070 | 0.12 |
| 8402-3 | 58 × 132 | 1.1 | 0.061 | 0.4 | 0.4 | 0.60 | 0.05 | 12.3 | 561.1 | 0.015 | 1.02 | 5.46 | 85.7 | 5.5 | 0.6095 | 0.14 |
| Estherville (type 3–4) | | | | | | | | | | | | | | | | |
| ES-1 | 90 | 1.8 | 0.086 | 0.4 | 0.2 | 0.45 | 0.105 | 4.32 | 208.9 | 0.022 | 1.06 | 7.09 | 89.0 | 7.1 | 0.6083 | 0.32 |
| ES-2 | 60 | 0.53 | 0.040 | 0.2 | 0.3 | 0.26 | 0.086 | 3.06 | 154.3 | 0.010 | 1.08 | 17.92 | 90.7 | 17.9 | 0.6110 | 0.49 |
| ES-3 | 130 | 5.4 | 0.107 | 1.2 | 0.2 | 1.78 | 0.108 | 16.4 | 740.4 | 0.027 | 1.01 | 3.66 | 85.1 | 3.7 | 0.60889 | 0.12 |
| ES-4 | 110 | 3.3 | 0.081 | 0.4 | 0.1 | 0.54 | 0.086 | 6.20 | 291.9 | 0.020 | 1.03 | 6.30 | 86.3 | 6.3 | 0.6088 | 0.24 |
| ES-5 | 50 | 0.32 | 0.065 | 0.4 | 1.2 | 0.46 | 0.088 | 5.19 | 247.1 | 0.017 | 1.00 | 7.22 | 83.8 | 7.2 | 0.6068 | 0.30 |
| ES-6 [k] | <40, 80 | | 0.126 | 0.2 | | 0.30 | 0.069 | 4.31 | 205.8 | 0.031 | 1.06 | 10.72 | 91.3 | 10.7 | 0.6221 | 0.33 |
| ES-7 | <40 | | 0.100 | 0.20 | | 0.05 | 0.060 | 0.91 | 60.1 | 0.030 | 0.93 | 50.52 | 77.5 | 50.5 | 0.6071 | 1.47 |
| ES-8 | <40 | | 0.016 | 0.04 | | 0.05 | 0.068 | 0.78 | 52.5 | 0.004 | 1.00 | 57.10 | 86.2 | 57.1 | 0.626 | 1.66 |
| NWA 8741 (type 4) | | | | | | | | | | | | | | | | |
| 8741-1 | 94 × 150 | 3.3 | 0.060 | 1.1 | 0.3 | 1.59 | 0.08 | 20.1 | 906.9 | 0.015 | 1.02 | 1.97 | 85.3 | 2.0 | 0.6076 | 0.10 |
| 8741-2 | 81 × 142 | 2.3 | 0.057 | 1.1 | 0.5 | 1.56 | 0.09 | 16.9 | 765.6 | 0.014 | 1.02 | 2.15 | 84.9 | 2.2 | 0.6072 | 0.10 |
| 8741-3 | 76 × 124 | 1.8 | 0.056 | 1.3 | 0.7 | 1.33 | 0.08 | 16.7 | 756.2 | 0.015 | 1.02 | 3.36 | 85.8 | 3.4 | 0.6085 | 0.12 |
| 8741-4 | 115 × 124 | 4.0 | 0.058 | 1.9 | 0.5 | 2.49 | 0.11 | 22.5 | 1014.0 | 0.014 | 1.01 | 1.20 | 84.5 | 1.2 | 0.6067 | 0.09 |
| 8741-5 | 110 × 154 | 4.6 | 0.059 | 4.0 | 0.9 | 5.67 | 0.10 | 54.7 | 2430.1 | 0.014 | 1.01 | 0.54 | 84.9 | 0.5 | 0.6078 | 0.05 |
| 8741-6 | 156 × 217 | 13 | 0.062 | 2.0 | 0.2 | 2.79 | 0.09 | 32.7 | 1462.3 | 0.015 | 1.02 | 1.07 | 84.9 | 1.1 | 0.6072 | 0.07 |

**Table S1b.** (*continued*)

| Sample | Dates (Ma) | | | | | | MSWD |
|---|---|---|---|---|---|---|---|
| | $^{206}Pb/^{238}Pb$ [j] | ± 2σ (abs) | $^{207}Pb/^{235}Pb$ [j] | ± 2σ (abs) | $^{207}Pb/^{206}Pb$ [j] | ± 2σ (abs) | |
| NWA 8402 (type 3) | | | | | | | |
| 8402-1 | 4485 | 78 | 4511 | 24 | 4523.1 | 2.0 | |
| 8402-2 | 4504 | 72 | 4518 | 22 | 4523.8 | 1.9 | |
| 8402-3 | 4534 | 178 | 4531 | 55 | 4529.8 | 2.1 | |
| | | | **Weighted mean** | | **4525.4** | **1.1** | **13** |
| Estherville (type 3–4) | | | | | | | |
| ES-1 | 4664 | 235 | 4569 | 71 | 4527.0 | 4.6 | |
| ES-2 | 4714 | 599 | 4588 | 180 | 4533.3 | 7.1 | |
| ES-3 | 4513 | 119 | 4524 | 37 | 4528.4 | 1.7 | |
| ES-4 | 4560 | 206 | 4538 | 63 | 4528.1 | 3.5 | |
| ES-5 | 4476 | 233 | 4509 | 72 | 4523.4 | 4.4 | |
| ES-6 [k] | 4673 | 356 | 4594 | 108 | 4559.5 | 4.9 | |
| ES-7 | 4225 | 1566 | 4430 | 507 | 4524 | 21 | |
| ES-8 | 4469 | 1841 | 4537 | 573 | 4568 | 24 | |
| | | | **Weighted mean without ES-6** | | **4527.9** | **1.4** | **1.4** |
| NWA 8741 (type 4) | | | | | | | |
| 8741-1 | 4527 | 64 | 4526 | 20 | 4525.3 | 1.5 | |
| 8741-2 | 4517 | 70 | 4522 | 22 | 4524.3 | 1.6 | |
| 8741-3 | 4543 | 110 | 4532 | 34 | 4527.4 | 1.9 | |
| 8741-4 | 4503 | 39 | 4517 | 12 | 4523.0 | 1.3 | |
| 8741-5 | 4511 | 17 | 4521 | 5 | 4525.8 | 0.8 | |
| 8741-6 | 4517 | 35 | 4522 | 11 | 4524.3 | 1.1 | |
| | | | **Weighted mean** | | **4525.0** | **0.5** | **4.4** |
| | | | **Weighted mean age of VM-5, ES-6, and ES-8** | | **4558.5** | **0.8** | **0.39** |
| | | | **Weighted mean age without VM-5, ES-6, and ES-8** | | **4525.4** | **0.4** | **3.8** |

**Table S2.** Previously reported $^{146,147}$Sm-$^{142,143}$Nd, $^{207}$Pb-$^{206}$Pb, and $^{244}$Pu-Xe dates of basaltic eucrites and $^{207}$Pb-$^{206}$Pb and $^{182}$Hf-$^{182}$W dates of eucritic zircons.

| Meteorite | Isotope system | Sample type/method | Date (Ma) | | | Histogram* | Reference |
|---|---|---|---|---|---|---|---|
| Basaltic eucrite | | | | | | | |
| Asuka 880702 | $^{244}$Pu-Xe | WR/noble gas MS | 4565 | ± | 15 | x | [73] |
| Asuka 880761 | $^{244}$Pu-Xe | WR/noble gas MS | 4538 | ± | 15 | x | [73] |
| Asuka 881388 | $^{244}$Pu-Xe | WR/noble gas MS | 4530 | ± | 32 | x | [73] |
| | $^{182}$Hf-$^{182}$W | zircon (n = 4)/SIMS | 4563.8 | ± | 3.8 | x | [27] |
| | $^{207}$Pb-$^{206}$Pb | zircon (n = 5)/SIMS | 4555 | ± | 54 | x | [23] |
| Asuka 881467 | $^{182}$Hf-$^{182}$W | zircon (n = 6)/SIMS | 4561.0 | ± | 4.6 | x | [27] |
| | $^{207}$Pb-$^{206}$Pb | zircon (n = 8)/SIMS | 4558 | ± | 13 | x | [23] |
| Agoult | $^{207}$Pb-$^{206}$Pb | plag (isochron)/TIMS | 4533.9 | ± | 0.7 | x | [74] |
| | $^{207}$Pb-$^{206}$Pb | zircon (n = 8)/TIMS | 4554.5 | ± | 2 | x | [25] |
| Béréba | $^{207}$Pb-$^{206}$Pb | plag, WR (isochron)/TIMS | 4522 | ± | 4 | x | [75] |
| | $^{244}$Pu-Xe | WR/noble gas MS | 4512 | ± | 18 | | [76] |
| | $^{244}$Pu-Xe | WR/noble gas MS | 4498 | ± | 16 | | [31] |
| | $^{244}$Pu-Xe | | 4504 | ± | 12 | x | weighted mean date of 4512 ± 18 and 4498 ± 16† |
| | $^{207}$Pb-$^{206}$Pb | zircon (n = 5)/SIMS | 4552 | ± | 20 | x | [24] |
| Binda | $^{244}$Pu-Xe | WR/noble gas MS | 4529 | ± | 34 | x | [76] |
| Bouvante | $^{244}$Pu-Xe | WR/noble gas MS | 4547 | ± | 15 | x | [31] |
| Cachari | $^{244}$Pu-Xe | WR/noble gas MS | ~4498 | | | | [31] |
| | $^{244}$Pu-Xe | glass/noble gas MS | ~4517 | | | | [31] |
| | $^{244}$Pu-Xe | | 4508 | | | | weighted mean date of 4498 Ma and 4517 Ma† |
| | $^{207}$Pb-$^{206}$Pb | zircon (n = 17)/SIMS | 4551 | ± | 14 | x | [24] |
| Caldera | $^{147}$Sm-$^{143}$Nd | plag, pyx, WR (isochron)/TIMS | 4544 | ± | 19 | x | [77] |
| | $^{207}$Pb-$^{206}$Pb | plag, pyx (isochron)/TIMS | 4516 | ± | 2.8 | x | [78] |
| | $^{244}$Pu-Xe | WR/noble gas MS | ~4513 | | | | [31] |
| | $^{207}$Pb-$^{206}$Pb | zircon (n = 4)/SIMS | 4542 | ± | 80 | x | [24] |
| Camel Donga | $^{207}$Pb-$^{206}$Pb | Pyx (model age)/TIMS | 4510.9 | ± | 1 | x | [79] |
| | $^{244}$Pu-Xe | WR/Noble gas MS | 4507 | ± | 16 | | [76] |
| | $^{244}$Pu-Xe | WR/Noble gas MS | 4521 | ± | 20 | | [31] |
| | $^{244}$Pu-Xe | | 4512 | ± | 12 | x | weighted mean date of 4507 ± 16 Ma and 4521 ± 20 Ma† |
| | $^{207}$Pb-$^{206}$Pb | zircon (n = 35)/SIMS | 4531 | ± | 10 | x | [24] |

**Table S2.** (*continued*)

| Meteorite | Isotope system | Sample type/method | Date (Ma) | | | Histogram* | Reference |
|---|---|---|---|---|---|---|---|
| Chervony Kut | $^{147}$Sm-$^{143}$Nd | plag, pyx, WR (isochron)/TIMS | 4580 | ± | 30 | | [80] |
| | $^{146}$Sm-$^{142}$Nd | plag, pyx, WR (isochron)/TIMS | 4550 | ± | 30 | x | [80] |
| | $^{207}$Pb-$^{206}$Pb | plag, pyx/TIMS | 4312.6 | ± | 1.6 | | [78] |
| | $^{244}$Pu-Xe | WR/Noble gas MS | 4538 | ± | 19 | x | [30] |
| Dar al Gani 380 | $^{207}$Pb-$^{206}$Pb | pyx (model age)/TIMS | 4527.1 | ± | 3.2 | x | [79] |
| Dhofar 182 | $^{182}$Hf-$^{182}$W | single zircon/SIMS | 4539 | ± | n.l.b./+11 | | [29] |
| | $^{182}$Hf-$^{182}$W | zircon (n = 4)/SIMS | 4563.3 | ± | 2.9 | x | [29] |
| Elephant Moraine 90020 | $^{147}$Sm-$^{143}$Nd | plag, pyx, WR (isochron)/TIMS | 4510 | ± | 40 | | [81] |
| | $^{146}$Sm-$^{142}$Nd | plag, pyx, WR (isochron)/TIMS | 4483 | ± | 26 | x | [81] |
| | $^{182}$Hf-$^{182}$W | single zircon/SIMS | <4533.3 | | | | [27] |
| Hammadah al Hamra 286 | $^{182}$Hf-$^{182}$W | zircon (n = 3)/SIMS | 4551.1 | ± | 8.1 | x | [29] |
| Jonzac | $^{244}$Pu-Xe | WR/noble gas MS | 4472 | ± | 16 | x | [31] |
| Juvinas | $^{147}$Sm-$^{143}$Nd | internal isochron/TIMS | 4560 | ± | 80 | x | [82] |
| | $^{207}$Pb-$^{206}$Pb | plag, pyx, glass, WR (isochron)/TIMS | 4518 | ± | 16 | x | [83] |
| | $^{207}$Pb-$^{206}$Pb | plag, pyx, glass, WR (isochron)/TIMS | 4571 | ± | 14 | x | [83] |
| | $^{207}$Pb-$^{206}$Pb | plag, pyx, WR (isochron)/TIMS | 4539 | ± | 4 | x | [84] |
| | $^{207}$Pb-$^{206}$Pb | plag, pyx (isochron)/TIMS | 4320.9 | ± | 1.7 | | [78] |
| | $^{244}$Pu-Xe | WR/noble gas MS | 4551 | ± | 15 | x | [31] |
| | $^{207}$Pb-$^{206}$Pb | zircon (n = 20)/SIMS | 4545 | ± | 15 | x | [24] |
| Millbillillie | $^{244}$Pu-Xe | WR/noble gas MS | 4566 | ± | 24 | x | [76] |
| | $^{244}$Pu-Xe | WR/noble gas MS | 4507 | ± | 21 | | [76] |
| | $^{244}$Pu-Xe | WR/noble gas MS | 4522 | ± | 16 | | [31] |
| | $^{244}$Pu-Xe | | 4516 | ± | 13 | x | weighted mean date of 4507 ± 21 Ma and 4522 ± 16 Ma† |
| | $^{207}$Pb-$^{206}$Pb | single zircon (core)/SIMS | 4555 | ± | 17 | | [26] |
| | $^{207}$Pb-$^{206}$Pb | single zircon (rim)/SIMS | 4531 | ± | 20 | | [26] |
| Nuevo Laredo | $^{244}$Pu-Xe | WR/noble gas MS | 4507 | ± | 27 | x | [31] |
| | $^{207}$Pb-$^{206}$Pb | pyx, plag, WR (isochron)/TIMS | 4534 | ± | 2 | x | [75] |
| Northwest Africa 049 | $^{207}$Pb-$^{206}$Pb | pyx (model age)/TIMS | 4439.1 | ± | 4.5 | x | [79] |
| Northwest Africa 1908 | $^{182}$Hf-$^{182}$W | single zircon/SIMS | 4557.3 | ± | 1.6 | | [29] |
| | $^{182}$Hf-$^{182}$W | single zircon/SIMS | 4564 | ± | -5/+3 | | [29] |

**Table S2.** (*continued*)

| Meteorite | Isotope system | Sample type/method | Date (Ma) | | | Histogram* | Reference |
|---|---|---|---|---|---|---|---|
| Northwest Africa 4523 | $^{182}$Hf-$^{182}$W | zircon (n = 5)/SIMS | 4561.5 | ± | 3.8 | x | [29] |
| Northwest Africa 5073 | $^{182}$Hf-$^{182}$W | single zircon/SIMS | 4532 | ± | -11/+6 | | [29] |
| | $^{182}$Hf-$^{182}$W | single zircon/SIMS | 4547 | ± | 2.7 | | [29] |
| | $^{182}$Hf-$^{182}$W | single zircon/SIMS | 4561 | ± | -7/+5 | | [29] |
| Northwest Africa 5356 | $^{182}$Hf-$^{182}$W | zircon (n = 7)/SIMS | 4562.9 | ± | 2.6 | x | [29] |
| Northwest Africa 7188 | $^{147}$Sm-$^{143}$Nd | plag, pyx, WR (isochron)/TIMS | 4582 | ± | 190 | | [85] |
| | $^{146}$Sm-$^{142}$Nd | plag, pyx, WR (isochron)/TIMS | 4553 | ± | -17/+20 | x | [85] |
| Padvarninkai | $^{244}$Pu-Xe | WR/noble gas MS | 4520 | ± | 22 | x | [31] |
| | $^{207}$Pb-$^{206}$Pb | zircon (n = 18)/SIMS | 4555 | ± | 13 | x | [23] |
| Petersburg | $^{244}$Pu-Xe | WR/noble gas MS | 4541 | ± | 25 | x | [31] |
| Piplia Kalan | $^{147}$Sm-$^{143}$Nd | plag, pyx, WR (isochron)/TIMS | 4570 | ± | 23 | x | [86] |
| Sioux County | $^{244}$Pu-Xe | WR/noble gas MS | 4499 | ± | 17 | x | [31] |
| Stannern | $^{147}$Sm-$^{143}$Nd | plag, pyx, WR (isochron)/TIMS | 4480 | ± | 70 | x | [87] |
| | $^{244}$Pu-Xe | WR/noble gas MS | 4434 | ± | 13 | x | [76] |
| | $^{207}$Pb-$^{206}$Pb | zircon (n = 4)/SIMS | 4550 | ± | 10 | x | [88] |
| Vetluga | $^{244}$Pu-Xe | WR/noble gas MS | 4479 | ± | 22 | x | [31] |
| Yamato 75011 | $^{147}$Sm-$^{143}$Nd | plag, pyx, WR (isochron)/TIMS | 4550 | ± | 140 | x | [89] |
| | $^{207}$Pb-$^{206}$Pb | zircon (n = 5)/SIMS | 4550.1 | ± | 9.4 | x | [23] |
| Yamato 791438 | $^{207}$Pb-$^{206}$Pb | zircon (n = 2)/SIMS | 4534 | ± | 20 | x | [28] |
| Yamato 792510 | $^{147}$Sm-$^{143}$Nd | plag, pyx, WR (isochron)/TIMS | 4570 | ± | 90 | x | [90] |
| | $^{207}$Pb-$^{206}$Pb | zircon (n = 13)/SIMS | 4545 | ± | 18 | x | [23] |

*Data marked x are shown in Fig. 2.

†Ages obtained from the same chronometer were combined if they are identical within the analytical uncertainties.

n.l.b. = no lower bound on date.

Other abbreviations are as follows: WR, whole rock; plag, plagioclase; pyx, pyroxene.

References are shown in the next sheet.

**REFERENCES**

59. Ewing, R. C., Meldrum, A., Wang, L. M., Weber, W. J. & Corrales, L. R. Radiation effects in zircon. *Rev. Mineral. Geochem.* **53**, 387–425 (2003).

60. Wiedenbeck, M. *et al.* Further characterization of the 91500 zircon crystal. *Geostand. Geoanal. Res.* **28**, 9–39 (2004).

61. Wiedenbeck, M. *et al.* Three natural zircon standards for U–Th–Pb, Lu–Hf, trace-element and REE analyses, *Geostand. Newsl.* **19**, 1 – 23 (1995).

62. Mattinson, J. M., Zircon U–Pb chemical abrasion (CA-TIMS) method: combined annealing and multi-step partial dissolution analysis for improved precision and accuracy of zircon ages. *Chem. Geol.* **220**, 47–66 (2005).

63. Delaney, J. S., Prinz, M. & Takeda, H. The polymict eucrites. *Proc. Lunar Planet. Sci. Conf. 15th, in J. Geophys. Res.*, **89**, C252–C288 (1984).

64. Yamaguchi, A. *et al.* Experimental evidence of fast transport of trace elements in planetary basaltic crusts by high temperature metamorphism. *Earth Planet. Sci. Lett.* **368**, 101–109 (2013).

65. Warren, P. H., Kallemeyn, G. W., Huber, H., Ulff-Møller, F. & Choe, W. Siderophile and other geochemical constraints on mixing relationships among HED-meteoritic breccias. *Geochim. Cosmochim. Acta* **73**, 5918–5943 (2009).

66. Kiesl, W., Weinke, H. H. & Wichtl, M. The Medanitos meteorite. *Meteoritics* **13,** 513–516 (1978).

67. Hsu, W. & Crozaz, G. Mineral chemistry and the petrogenesis of eucrites: II. Cumulate eucrites. *Geochim. Cosmochim. Acta* **61**, 1293–1302 (1997).

68. Crozaz, G. & Wadhwa, M. The terrestrial alteration of Saharan Shergottites Dar al Gani 476 and 489: A case study of weathering in a hot desert environment. *Geochim. Cosmochim. Acta* **65**, 971–978 (2001).

69. Cook, D. L., Walker, R. J., Horan, M. F., Wasson, J. T. & Morgan, J. W. Pt-Re-Os systematics of group IIAB and IIIAB iron meteorites. *Geochim. Cosmochim. Acta* **68***, 1413–1431 (2004).

70. Smoliar, M. I., Walker, R. J. & Morgan, J. W. Re-Os ages of group IIA, IIIA, IVA, and IVB iron meteorites. *Science*, **271**, 1099–1102 (1996).

71. Hunt, A. C. *et al*. Late metal–silicate separation on the IAB parent asteroid: constraints from combined W and Pt isotopes and thermal modelling. *Earth Planet. Sci. Lett.* **482**, 490–500 (2018).

72. Neumann, W., Breuer, D. & Spohn, T. Differentiation of Vesta: implications for a shallow magma ocean. *Earth Planet. Sci. Lett*. **395**, 267–280 (2014).

73. Park, J. & Nagao, K. Noble gas and chronological study of Asuka eucrites: A-880761 and A-881388 are paired, but A-880702 is not. *Antarct. Meteorite Res*. **18**, 213–224 (2005).

74. Iizuka, T. et al. Thermal history of basaltic eucrites as recorded by lead and argon isotopes. *49th Lunar and Planetary Science Conference*, abstr. No. 2083 (2018).

75. Carlson, R. W., Tera, F. & Boctor, N. Z. Radiometric geochronology of the eucrites Nuevo Laredo and Bereba. *Lunar and Planetary Science Conference* **19**, abstr. 166 (1988).

76. Miura, Y. N., Nagao, K., Sugiura, N., Fujitani, T. & Warren, P. H. Noble gases, 81Kr-Kr exposure ages, and 244Pu-Xe ages of six eucrites, Béréba, Binda, Camel Donga, Juvinas, Millbillillie, and Stannern. *Geochim. Cosmochim. Acta* **62**, 2369–2388 (1998).

77. Wadhwa, M. & Lugmair, G. W. Age of the eucrite Caldera from convergence of long-lived and short-lived chronometers. *Geochim. Cosmochim. Acta* **60**, 4889–4893.(1996)

78. Galer, S. J. G. & Lugmair, G. W. Lead isotope systematics of noncumulate eucrites. *Meteoritics & Planetary Science*, **31**, abstr. A47 (1996).

79. Iizuka, T., Kaltenbach, A., Amelin, Y., Stirling, C. H. & Yamaguchi, A. U-Pb isotope systematics of eucrites in relation to their thermal history. *44th Lunar and Planetary Science Conference*, abstr. No. 1719 (2013).

80. Wadhwa, M. & Lugmair, G. W. Sm-Nd systematics of the eucrite Chervony Kut. *Lunar and Planetary Science Conference*, **26**, abstr. 1453 (1995).

81. Yamaguchi, A. et al. Post-crystallization reheating and partial melting of eucrite EET90020 by impact into the hot crust of asteroid 4 Vesta 4.50 Ga ago. *Geochim. Cosmochim. Acta* **65**, 3577–3599 (2001).

82. Lugmair, G. W. Sm-Nd ages: a new dating method. Meteoritics, 9, abstr. 369 (1974).

83. Tatsumoto, M. & Unruh, D. M. Formation and early brecciation of the Juvinas achondrite inferred from U-Th-Pb systematics. *Meteoritics*, **10**, abstr. 500 (1975).

84. Manhes, G., Allegre, C. J. & Provost, A. U-Th-Pb systematics of the eucrite Juvinas: Precise age determination and evidence for exotic lead. *Geochim. Cosmochim. Acta* **48**, 2247–2264 (1984).

85. Kagami, S., Haba, M. K., Yokoyama, T., Usui, T. & Greenwood, R. C. Geochemistry, petrology, and Sm-Nd dating of a Stannern group eucrite, Northwest Africa 7188, *49th Lunar and Planetary Science Conference*, abstr. No. 2083 (2018).

86. Kumar, A., Gopalan, K. & Bhandari, N. 147Sm-143Nd and 87Rb-87Sr ages of the eucrite Piplia Kalan. *Geochim. Cosmochim. Acta* **63**, 3997–4001 (1999).

87. Lugmair, G. W. & Scheinin, N. B. Sm-Nd systematics of the Stannern meteorite. *Meteoritics*, **10**, abstr. 447–448 (1975).

88. Ireland, T. R. & Bukovanska, M. Zircons from the Stannern eucrite. *Meteoritics*, **27**, abstr. 237 (1992).

89. Nyquist, L. E. et al. Rb-Sr and Sm-Nd internal isochron ages of a subophitic basalt clast and a matrix sample from the Y75011 eucrite. *Journal of Geophysical Research* **91**, 8137–8150 (1986).

90. Nyquist, L. E. et al. Crystallization, recrystallization and impact metamorphic ages of eucrites Y-792510 and Y-791186. *Geochim. Cosmochim. Acta* **61**, 2119–2138 (1997).